\documentclass{aa}  

\usepackage{graphicx}
\usepackage{txfonts}
\usepackage{multirow}
\usepackage{longtable,lscape}
\usepackage{color}
\usepackage{multirow}
\usepackage{ccaption}

\definecolor{rojo}{rgb}{1,0,0}
\definecolor{blu}{rgb}{0,0,1}

\begin{document}

\title{Three open clusters containing Cepheids: NGC\,6649, NGC\,6664 $\&$ Berkeley\,55}
\titlerunning{Three open clusters with Cepheids}

\author{
J. Alonso-Santiago\inst{1,2}
\and I. Negueruela\inst{3}
\and A. Marco\inst{2}
\and H. M. Tabernero\inst{2,4}
\and N. Castro\inst{5}
}

\institute{
INAF, Osservatorio Astrofisico di Catania, via S. Sofia 78, 95123, Catania, Italy\\
\email{javier.alonso@inaf.it} \\
\and Dpto de F\'{i}sica, Ingenier\'{i}a de Sistemas y Teor\'{i}a de la Se\~{n}al, Escuela Polit\'ecnica Superior, Universidad de Alicante, Carretera de San Vicente del Raspeig s/n, 03690, Spain \\
\and Dpto de F\'{i}sica Aplicada. Universidad de Alicante, Carretera de San Vicente del Raspeig s/n, 03690, Spain \\
\and Instituto de Astrof{\'i}sica e Ci{\^e}ncias do Espa\c{c}o,
Universidade do Porto, CAUP, Rua das Estrelas, 4150-762 Porto, Portugal \\
\and Leibniz-Institut für Astrophysik Potsdam, An der Sternwarte 16, 14482 Potsdam, Germany
}

\date{}

 
  \abstract
  {Classical Cepheids in open clusters play an important role in benchmarking stellar evolution models, anchoring the cosmic distance scale, and invariably
  securing the Hubble constant. NGC\,6649, NGC\,6664 and Berkeley\,55 are three pertinent clusters that host classical Cepheids and red (super)giants, and an 
  analysis was consequently initiated to assess newly acquired spectra ($\approx$50), archival photometry, and $Gaia$ DR2 data. Importantly, for the first time chemical 
  abundances are determined for the evolved members of NGC\,6649 and NGC\,6664. We find that they are slightly metal-poor relative to the mean Galactic gradient, and 
  an overabundance of Ba is observed. Those clusters likely belong to the thin disc, and the latter finding supports \citet{dor09} ``$s$-enhanced'' scenario. NGC\,6664 
  and Berkeley\,55 exhibit radial velocities consistent with Galactic rotation, while NGC\,6649 displays a peculiar velocity. The resulting age estimates for the clusters
  ($\approx$\,70\,Ma) imply masses for the (super)giant demographic of $\approx$\,6\,M$_{\sun}$. Lastly, the observed yellow-to-red (super)giant ratio is lower than expected, and 
  the overall differences relative to models reflect outstanding theoretical uncertainties.}


   \maketitle
%

\section{Introduction}\label{intro}

According to canonical models \citep[e.g.][]{Chiosi92}, stars spend most of their lives burning, in a stable way, hydrogen in their cores. Once the hydrogen fuel runs out, stars leave 
the main sequence (MS) expanding their outer envelopes and evolving towards lower effective temperatures. Depending on their masses they will become red giants (RGs) or supergiants (RSGs).
Either before they reach this stage (during the H-shell burning) or somewhat after (in the He-core burning), stars eventually turn into yellow supergiants (YSGs).
The ratio of stars observed in each of these brief phases of the life of a star, the yellow-to-red ratio, is a fundamental observable to constrain theoretical models. Specifically, 
it is key to get a better understanding of the so-called blue loop \citep{mat82,eks12,and14,wal15}.

The best places to test evolutionary models are star clusters. In them, we can calculate the ratios between the different evolved stars that they host.
More concretely, we focus on young open clusters to explore the mass boundary between RGs and RSGs \citep{ignacio17,salamanca}. Despite their physical and, hence, evolutionary 
differences, their morphological separation (in regards to the MK spectral classification) is not clear. Luminosity class I is expected for RSGs whereas class
III it is for RGs. However, the most massive intermediate-mass stars, with ages below 100\,Ma, are observed as K-type Ib supergiants \citep[see discussion in][]{Be55}.
Following our ongoing survey \citep{ignacio17,salamanca} with the aim of shedding some light on the issue, in this occasion we investigate three little-studied young open clusters, 
namely NGC\,6649, NGC\,6664 and Berkeley\,55, 
which contain a Cepheid (i.e. a YSG) each and some red GK (super)giants.

Beyond the evolutionary context, classical Cepheids (CCs) play a fundamental role in one of the hottest topics in modern astrophysics: the accurate determination of the present rate 
of the expansion of the Universe (the Hubble constant, $H_{\textrm{0}}$). As well-known standard candles, they are the first rung in the cosmic distance ladder. Galactic CCs are the cornerstone 
to calibrate, via the Period-Luminosity relationship (PLR), the distance to extragalactic CCs  and, in turn, nearby Type Ia supernovae. In this way, the SHoES Team, 
by calibrating the distance ladder with detached eclipsing binaries in the Large Magellanic Cloud (LMC), masers in M\,106 and Milky Way parallaxes, 
obtained $H_{\textrm{0}}$=74.0$\pm$1.4\,km\,s$^{-1}$\,Mpc$^{-1}$ \citep{Riess19b}. This value differs at the 4.4\,$\sigma$ level from the Planck result, $H_{\textrm{0}}$=67.4$\pm$0.5\,km\,s$^{-1}$\,Mpc$^{-1}$ 
\citep{Planck}, based on the cosmic microwave background (CMB) within the flat $\Lambda$ cold dark matter cosmological model workframe.
These signficant discrepancies, which arise when $H_{\textrm{0}}$ is inferred from direct (local) late-time measurements or early-Universe estimates based on cosmological models,   
are the so-called $H_{\textrm{0}}$ tension \citep[see][for a review]{Riess19a, Verde19}.
Nonetheless, there is still no consensus on this as one might think from the recent results of the Chicago-Carnegie Hubble Project \citep{Freedman20}.
They obtained, by calibrating the Tip Red Giant Branch in the LMC, a value for $H_{\textrm{0}}$ that is not significantly offset relative to the 
CMB ($H_{\textrm{0}}$=69.9$\pm$0.8\,km\,s$^{-1}$\,Mpc$^{-1}$). 

The existence of this tension, if real, could imply new physics beyond the standard cosmology \citep{Vagnozzi20}. However, it remains an open question in which the role of 
Cepheids might be key to relieve it. On the one hand, the impact of the blending caused by unresolved sources in the vicinity of extragalactic Cepheids
is currently a controversial issue. An appropriate blending correction, consistently applied by diverse groups, is necessary to assure a comparison of the different $H_{\textrm{0}}$
values on a homogeneous basis, which would surely lead to mitigate the tension \citep{Majaess20}. 
On the contrary, according to \citet{Riess20} the influence of systematic errors in the distant Cepheids backgrounds does not significantly affect it.
On the other hand, Galactic Cepheids are fundamental to calibrate the PLR since their distances can be accurately determined from their parallaxes, specially in the 
era of $Gaia$. The new parallaxes measured by it in conjunction with the existence of an underdensity in the local galaxy distribution (the ``Local Hole'')
might explain the $H_{\textrm{0}}$ tension \citep{Shanks19}, although about it there are also some discrepancies \citep{Riess18}. However, the current data release \citep[DR2,][]{GaiaDR2}, 
because of the uncertainty on the parallax zero-point offset, does not allow us to improve the PLR calibration and, thus, the distance scale \citep{Groenewegen18}.
In this context, Cepheids in open clusters are of great importance since they offer us an additional, and parallax-independent, way to estimate their distance: the MS fitting method.
In this paper we aim to explore this via from a comprehensive study of the clusters aforementioned (NGC\,6649, NGC\,6664 and Berkeley\,55) which host the 
Cepheids V367\,Sct, EV\,Sct and the little-known Be55$\#$107, respectively.

\section{Target clusters}

A brief description of the properties for the clusters studied in this work is summarised below:

\subsection{NGC\,6649}

The first of the targets is a compact heavily-reddened cluster located in the first Galactic quadrant [$\alpha$(2000)\,=\,18h\:33m\:27s, $\delta$(2000)\,=\,$-10^{\circ}\:24\arcmin\:12\arcsec$; $\ell=21\fdg635$, $b=-0\fdg785$].
It is well-known for hosting the double-mode Cepheid V367\,Sct, whose cluster membership was suggested for the first time by \citet{Ro63}. 
Among the first photometric studies carried out on this cluster, the most complete is that performed by \citet{Ma75}. They obtained photoelectric photometry for 68 stars and photographic data for more than 400 stars. They derived a 
reddening of $E(B-V)$=1.37$\pm$0.01 and placed the cluster at a distance of 1.7$\pm$0.6\,kpc. By taking some spectra they also noticed the presence of three emission-line stars. 
\citet{Turner81} re-examined the previous photometry and derived an age for the cluster around 50\,Ma.
\citet{Wa87} carried out the first modern $UBV$ CCD photometry for 400 stars. They derredened individually their observations but without quoting the reddening found. Their estimate for distance is 1.6$\pm$0.1\,kpc, pretty similar to the previous
from \citet{Ma75}. Considering the period of the Cepheid (log\,$P$<1), they assumed for NGC\,6649 an age similar to the Pleiades.
More recently \citet{Hoy03}, also by employing CCD photometry, derived for the cluster a mean reddening of $E(B-V)$=1.37$\pm$0.06 and a distance of 1.8$\pm$0.1\,kpc. Further studies, beyond characterising the Cepheid are
scarce. \citet{Me08} obtained the average radial velocity (RV) for the cluster from four evolved stars and \citet{be_2345} noted the high fraction of Be stars in this cluster. Beyond these papers, there are no other spectroscopic studies and the cluster age 
still remains somewhat uncertain.

\subsection{NGC\,6664}\label{intro_6664}

It is a young open cluster located very close to NGC\,6649 [$\alpha$(2000)\,=\,18h\:36m\:42s, $\delta$(2000)\,=\,$-08^{\circ}\:13\arcmin\:00\arcsec$; $\ell=23\fdg945$, $b=-0\fdg491$]\footnote
{Equatorial coordinates from the WEBDA database. Simbad/Aladin provide wrong values for the cluster, $\alpha$(2000)\,=\,18h\:36m\:37s, $\delta$(2000)=\,$-07^{\circ}\:48\arcmin\:48\arcsec$.}.
It is a little-studied cluster, best-known for hosting the Cepheid EV\,Sct, whose possible cluster membership was firstly claimed by \citet{Kra57}. With the aim of confirming it, \citet{Arp58} and \citet{kr58}
performed the first studies of this cluster. \citet{Arp58} obtained $UBV$ photoelectric and photographic photometry for 31 and 29 stars, respectively. \citet{kr58} complemented the previous work providing
spectral classification for 11 stars and RVs for nine, the Cepheid included. These authors found an average reddening $E(B-V)$=0.60 mag and placed the cluster at 1.5\,kpc. They obtained a mean
radial velocity, $v_{\textrm{rad}}=23\pm$2\,km\,s$^{-1}$, compatible with that of the Cepheid. From the absolute magnitude of the brightest MS stars of the cluster they estimated an
age slightly older than the Pleiades.
Later, \citet{scm6664} carried out $ubvy\beta$ photoelectric photometry on this cluster for 15 stars among the brightest stars from \citet{Arp58}. He derived a mean colour excess for the cluster 
$E(b-y)=0.56$, equivalent to $E(B-V)$=0.75, somewhat higher than that found by \citet{Arp58}. He obtained a distance of 1.38$\pm$0.06\,kpc and an age slightly younger than the Pleiades. More recently, \citet{Hoy03} carried out $UBV$ 
CCD photometry of clusters containing Cepheids. For NGC\,6664 they found $E(B-V)=0.66\pm0.06$ and $d=1.4\pm0.1$\,kpc, values ranging between those previously cited.
Finally, \citet{Me08} obtained for the cluster a mean $v_{\textrm{rad}}=18.6\pm0.6$\,km\,s$^{-1}$ from red giants members. They also suggested the binarity of some of the giants.
In addition, also the Cepheid EV\,Sct, because of the line profile variations observed in its spectrum, initially seemed to be part of a binary system \citep{kov99}. However, a deeper subsequent analysis 
suggested that a single object pulsating in a non-radial mode better explained its spectral features \citep{kov03}.

\subsection{Berkeley\,55}

The third cluster under study is Berkeley\,55 (Be\,55), placed in the second Galactic quadrant [$\alpha$(2000)\,=\,21h\:16m\:58s, $\delta$(2000)\,=\,$+51^{\circ}\:45\arcmin\:32\arcsec$; $\ell=93\fdg027$, $b=+1\fdg798$]. 
Be\,55 is a compact and reddened cluster as noted by \cite{Maciejewski07} who estimated the cluster parameters from $BV$ photometry: 
$E(B-V)$=1.74, $r_c$=0.7$\arcmin$, $\tau$=315\,Ma and a $d$=1.21\,kpc. \cite{Tadross08} by employing proper motions from the NOMAD catalogue and 2MASS photometry 
found a lower reddening ($E(B-V)$=1.5), a similar age ($\tau$=300\,Ma) and a slightly further distance ($d$=1.44\,kpc).
\cite{Be55} performed the only study devoted exclusively to this cluster, including the first spectroscopic observations as well as $UBV$ and 2MASS photometry.  
They found a significant evolved population composed of one yellow supergiant and six red (super)giants. They computed a reddening, $E(B-V)$=1.85 and placed 
the cluster roughly at 4\,kpc. The main difference with respect to previous papers, consequence of the different computed distance, lies in the age of the cluster, 50 Ma, a value considerably younger.
However, this age is fully consistent with the brightest blue members, mid-B giants.
More recently, \cite{Molina18} by using optical and infrared photometry together with low-resolution spectroscopy derived values for the cluster parameters 
compatible with the results of \cite{Be55}, confirming the younger age for the cluster, in the 30--100\,Ma range. 
\cite{Lohr18} identified the yellow supergiant present in the cluster as a Cepheid. 

\section{Observations}

\subsection{Spectroscopy}
We collected spectra in the field of the three clusters aforementioned for 44 stars in different runs which are described below. Our targets were selected among 
the cluster members according to the literature, specifically the brightest blue members (at the upper MS) and the evolved members. Table\,\ref{targets} summarises for these stars 
the cluster field in which they have been observed, the spectrograph used, properties of each spectrum such as the exposure time ($t_{\textrm{exp}}$) and the signal-to-noise ratio ($S/N$), and
the spectral type assigned.

Thirty-three stars were observed, in four runs, with ISIS which is mounted at the Cassegrain focus of the 4.2-m William Herschel Telescope (WHT), located at El Roque 
de los Muchachos Observatory in La Palma, in the Canary Islands (Spain). It is a high-efficiency, double-armed, medium-resolution
spectrograph, capable of long-slit work up to $\approx4'$ slit length and $\approx22''$ slit width. Because of the use of dichroic filters, simultaneous observing in both arms is possible.
ISIS is equipped with two 4k\,x\,2k CCDs, the thinned and blue-sensitive EEV12 (13.5 micron) on the blue arm and RED+ (15 micron) on the red one, a red-sensitive with almost
no fringing camera. We took spectra for 10 stars in the field of NGC\,6649 on May 20 and June 2, 2004. We used the grating R300B and a 
slit with a width of 1$\farcs$5, providing a low resolving power, $R\,\approx$\,730. On July 25, 2009 we observed the field of NGC\,6664 obtaining spectra for 11 stars. 
In this occasion we employed the same grating but with a slit width of 0$\farcs$98 at a resolution slightly higher, $R\,\approx$\,1\,100.
The last run was carried out on July 26, 2011. This time 12 stars were observed in the field of Be\,55 by employing the gratings R300B and R600R together with a slit with a 
width of 1$\farcs$5. These spectra cover a wavelength in the 2\,720--6\,340\,\AA{} interval for the blue arm and 7\,530--9\,130\,\AA{} for the red arm (with a $R\,\approx$\,3\,000).
The spectra were reduced following standard procedures with the {\scshape starlink} software. 

In the field of NGC\,6649, high-resolution spectra for six stars were taken with FEROS in May 2015 during the nights of 29th and 30th under ESO programme 095.A-9020(A).
FEROS \citep{feros} is mounted on the ESO/MPG 2.2-m telescope at La Silla Observatory (Chile) and is fibre-fed from the Cassegrain 
focus. It is an \'{e}chelle spectrograph which covers, in 39 orders, the wavelength range from 3\,500\,\AA{} to 9\,200\,\AA{}, providing a resolving power of $R=48\,000$. 
It has as a detector an EEV 2k\,x\,4k CCD and is fed by two fibres that provide simultaneous spectra of the astrophysical target plus either sky or one of the two calibration 
lamps. The reduction process was carried out by employing the {\sc{feros-drs}} pipeline based on {\sc{midas}} routines. 

We observed the five evolved stars of NGC\,6664 with HERMES \citep{hermes} during 2016 September 26--30. It is mounted on the 1.2-m Mercator telescope (La Palma). 
It is fibre-fed from the Nasmyth A focus, but it is situated in a separate room whose temperature is controlled. HERMES is an \'{e}chelle spectrograph which covers in a single 
exposure the wavelength range from 3\,760\,\AA{} to 9\,000\,\AA{} in 55 orders. Two observing modes are available: the high resolution mode (HRF) and the simultaneous wavelength 
reference mode (LRF). We used the former, which provides a resolving power, $R=85\,000$. HERMES is equipped with an E2V\,42-90 detector of 2k\,x\,4k pixels, with a wavelength-dependent
anti-reflective coating that greatly reduces fringing. Raw spectra were automatically reduced using the corresponding pipeline, {\scshape hermes-drs}, depending on {\scshape python} routines.

Finally, we took spectra of the red stars in Be\,55 with the Intermediate Dispersion Spectrograph (IDS) on the nights of 2017 September 24 and 26. IDS is mounted 
on the 2.5-m Isaac Newton Telescope (INT) at El Roque de los Muchachos (La Palma). We used the grating H1800V together a 1.5\arcsec slit, configuration which provided us a resolution, $R\approx9\,000$.
The setup is centred on 6\,700\,\AA{} covering a wavelength range from 6\,200\,\AA{} to 7\,200\,\AA{}. 
These spectra were reduced by employing the {\scshape iraf}\footnote{{\scshape iraf}
is distributed by the National Optical Astronomy Observatories, which are operated by the Association of Universities for Research in Astronomy, Inc., under the cooperative agreement
with the National Science Foundation.} packages following standard procedures. 

\begin{table}[h!]
  \caption{Stars observed spectroscopically in this work}
\begin{center}
\begin{tabular}{lcccc}
\hline\hline
Star & Sp T & $t_{\textrm{exp}}$ (s) & $S/N$ & Spectrograph \\
\hline
\multicolumn{5}{c}{NGC\,6649}\\
\hline
9            &  B1\,IIIe  & 1\,000 & 90  &  ISIS \\
14           &  B5\,V     & 1\,000 & 90  &  ISIS \\
19           &  F5\,Ib-II & 1\,000 & 90  &  ISIS \\
23           &  B7\,III   &  750   & 100 &  ISIS \\
28           &  B4\,III   &  600   & 100 &  ISIS \\
33           &  B6\,IV    &  750   & 90  &  ISIS \\
35           &  B5\,V     &  750   & 90  &  ISIS \\
42\,B        &  B8\,IV    & 1\,200 & 80  & FEROS \\
42\,K        &  K0\,Ib-II & 1\,200 & 80  & FEROS \\
48           &  B5\,V     & 1\,000 & 80  &  ISIS \\
49           &  K1\,Ib    & 1\,800 & 70  & FEROS \\
52           &  B5\,V    &  750   & 90  &  ISIS \\
58           &  B8\,II    &  600   & 120 &  ISIS \\
64$^{*}$     &  F7\,Ib    & 3\,000 & 90  & FEROS \\
111          &  M5\,Ib    & 3\,000 & 60  & FEROS \\
117          &  K5\,Ib    & 1\,200 & 70  & FEROS \\
\hline
\multicolumn{5}{c}{NGC\,6664}\\
\hline
50           & B9\,IV  &    300 & 120 &   ISIS  \\
51           & K0\,Ib  & 1\,500 & 50  &  HERMES \\
52           & G8\,Ib  & 2\,400 & 60  &  HERMES \\
53           & K2\,Ib  & 1\,800 & 50  &  HERMES \\
54           & K4\,Ib  & 2\,000 & 50  &  HERMES \\
55           & BN2\,IV &    200 & 120 &   ISIS  \\
56           & B6\,IIIe&    300 & 160 &   ISIS  \\
59           & B2.5\,V &    600 & 180 &   ISIS  \\
60           & B6\,IV  &    450 & 120 &   ISIS  \\
61           & B3\,IV  &    450 & 120 &   ISIS  \\
62           & B6\,IV  &    450 & 140 &   ISIS  \\
63           & B5\,III &    500 & 140 &   ISIS  \\
64           & B6\,V   &    500 & 110 &   ISIS  \\
80$^{*}$     & F8\,Ib  & 1\,800 & 50  &  HERMES \\
228          & B5\,Ve  & 1\,050 & 130 &   ISIS  \\
A            & B6\,IV  &    600 & 140 &   ISIS  \\
\hline 
\multicolumn{5}{c}{Berkeley\,55}\\
\hline
1            & K1\,Ib    & 4\,600 &  80  &   IDS \\ 
2            & K0\,Ib    & 3\,600 &  90  &   IDS \\ 
3            & G8\,II    & 3\,600 &  80  &   IDS \\ 
4            & K0\,Ib-II & 3\,600 &  60  &   IDS \\ 
5$^{*}$      & F8\,Ib    & 3\,600 & 100  &   IDS \\ 
6            & K4\,II    & 3\,600 &  90  &   IDS \\ 
7            & B4\,IV    & 3\,600 & 160  &   ISIS \\ 
10           & B6\,IV    & 3\,600 &  90  &   ISIS \\ 
11           & B6\,IV    & 3\,600 & 110  &   ISIS \\
12           & B6\,IV    & 4\,000 & 110  &   ISIS \\ 
17           & B5\,V     & 3\,600 & 110  &   ISIS \\ 
61           & M2\,II    & 3\,100 &  90  &   IDS \\ 
\hline
\end{tabular}
\end{center}
\begin{list}{}{}
\item[]$^{*}$ These stars are the Cepheids V367\,Sct (NGC\,6649), EV\,Sct (NGC\,6664) and Be\,55\,$\#$107, respectively.
  \end{list}
\label{targets}
\end{table}

\subsection{Archival data}
In order to carry out an analysis as accurate as possible, our spectroscopic observations are complemented with archival data. 
On the one hand, we resorted to photometric data found in the literature for our clusters. On the other hand, we took advantage of
all-sky surveys such as 2MASS and $Gaia$. For each cluster, we selected those sources inside a radius of $30\arcmin$ around the nominal cluster centre.
We only took the magnitudes of the $JHK_{\textrm{S}}$ from the 2MASS catalogue \citep{2MASS} for stars that have good-quality photometry (i.e. those 
without any `$U$' photometric flags in the catalogue). Additionally, we also employed both the photometric and the astrometric values provided in the $Gaia$ 
second data release \citep[DR2,][]{GaiaDR2} for those stars with sufficiently good astrometry (i.e. a parallax error smaller than 0.5 mas).

\section{Results}

Throughout this article, with the aim of identifying the stars observed, we followed the WEBDA numbering for NGC\,6649 and NGC\,6664. For S42 (NGC\,6649), a
visual double star (i. e. two stars very close on the sky without any physical relationship) not resolved by this designation, we added, behind the number, 
the letter corresponding to the spectral type of the star. In addition, we named with the letter ``A'' one star without a WEBDA numbering in NGC\,6664. 
In the case of Be\,55 we continue to use the numbering previously employed in \citet{Be55}.
All these stars are displayed in Table~\ref{targets} and in the finding charts which are shown for each cluster in the Appendix: Fig.~\ref{carta_6649} (NGC\,6649),
\ref{carta_6664} (NGC\,6664) and \ref{carta_be55} (Be\,55).

\subsection{Spectral classification}

We carried out the spectral classification of our targets following the same classical criteria used in previous works. A general description is provided below, but for further details see e.g. \citet{2345} and references therein. For our classification (in Table~\ref{targets}) we estimate a typical uncertainty
of around one spectral subtype. 

Our sample is composed mostly of blue stars covering the entire B spectral type. These stars have been observed with ISIS at low resolution mainly
for classification purposes, although we have also given an estimate of their atmospheric parameters. For these stars, we focused on the optical wavelength range
(4\,000\,--\,5\,000\,\AA), according to the specifications described by \citet{Ja87} and \citet{gray}. In this range some line ratios as \ion{Mg}{ii}\,
$\lambda$4481/\ion{He}{i}\,$\lambda$4471, \ion{Si}{ii}\,$\lambda$4128\,--\,30/\ion{Si}{iii}\,$\lambda$4553, \ion{Si}{ii}\,$\lambda$4128\,--\,30/\ion{He}{i}\,
$\lambda$4121, \ion{N}{ii}\,$\lambda$3995/\ion{He}{i}\,$\lambda$4009, and \ion{He}{i}\,$\lambda$4121/\ion{He}{i}\,$\lambda$4144 have been very useful.
On the other hand, we also have cool (super)giants with spectral types FGKM. In this case, these objects were observed at high resolution with FEROS and HERMES
with the aim of performing a detailed spectroscopic analysis. For classification, we payed attention to the near-infrared wavelengths around the \ion{Ca}{ii} 
triplet (8480--8750\,\AA{}). In this region many features of \ion{Fe}{i} (i.e. lines at 8514, 8621 and 8688\,\AA{}) and \ion{Ti}{i} (8518\,\AA{}) become stronger 
with later spectral types \citep{carquillat}. Additionally, some line ratios as \ion{Ti}{i}\,$\lambda$8518\,/\,\ion{Fe}{i}\,$\lambda$8514 and 
\ion{Ti}{i}\,$\lambda$8734\,/\,\ion{Mg}{i}\,$\lambda$8736  become larger with increasing spectral type, being very helpful for the classification.

\begin{figure*}  
  \centering         
  \includegraphics[width=16cm]{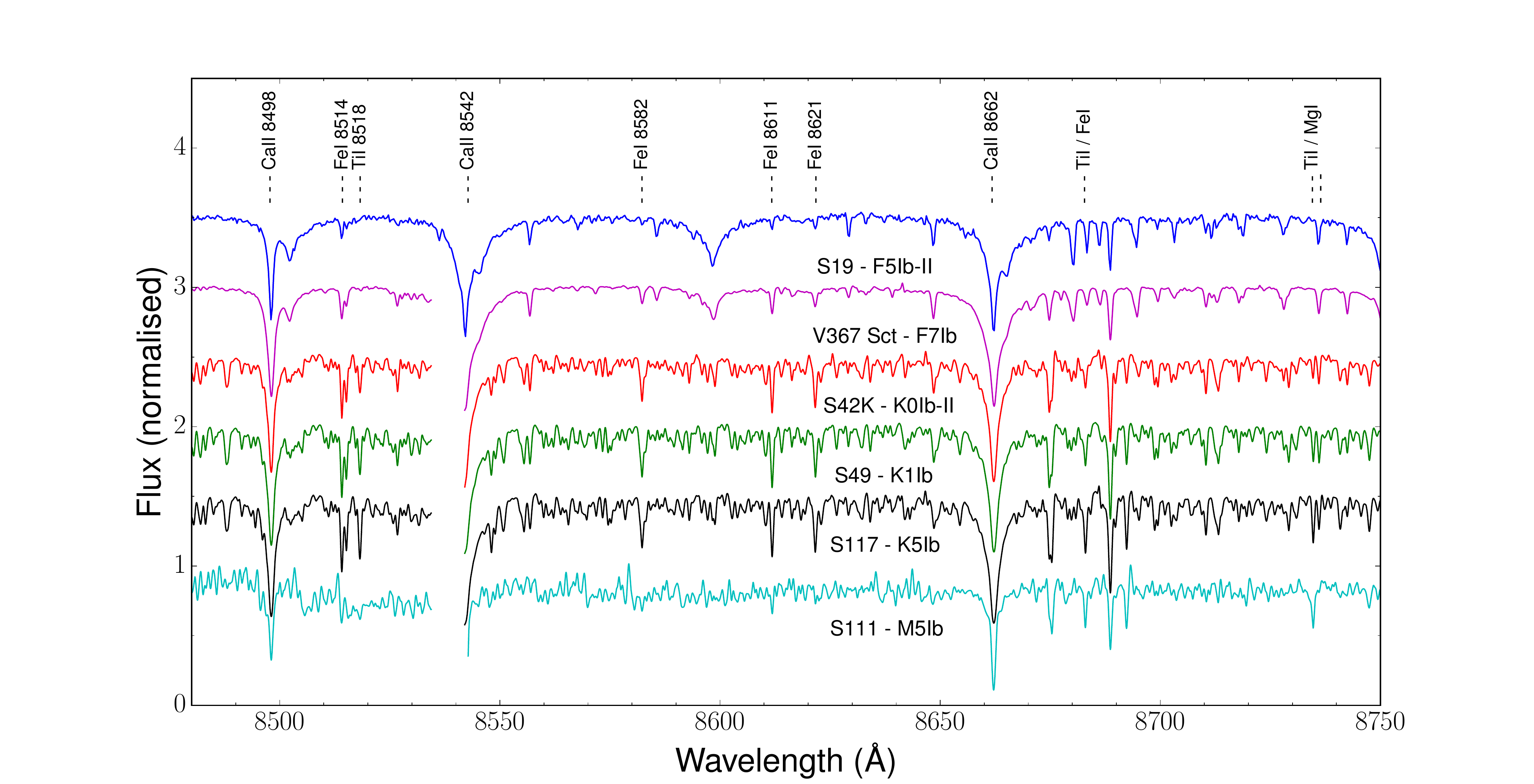}   
  \caption{Spectra of the cool stars observed in the field of NGC\,6649. With the only exception of S19, observed with ISIS, these stars
  were observed with FEROS (Note the gap between orders 37 and 38 around 8540\,\AA).} 
  \label{Rojas_6649}  
\end{figure*}

\textit{NGC\,6649:} In the field of this cluster we took spectra for 16 stars. Among them, we find 10 B-type stars almost covering the whole B spectral range. 
We also found that S9 is an extreme emission-line star (Be), showing $EW$(H$\beta$) of $-3$\,\AA{}. Besides this star, \citet{Ma75} found two other Be objects:
S28 and S58. We do not appreciate any emission in them around H$\beta$; it might be seen in H$\alpha$ but our spectra do not cover that spectral region.
\citet{be_2345} and \citet{mathew11} reported seven Be stars in this cluster, among which neither S28 nor S58 are found.
The only star in common with us is S9; we did not observed the other Be stars found in their works.
Regarding evolved stars, we observed six objects: two YSGs (S19 and the Cepheid V367\,Sct) and four RSGs: three K supergiants (S42\,K, S49, S117) and a cool M star (S111).
In Fig.\ref{Rojas_6649} the spectra of these cool stars are displayed.

\begin{figure*}  
  \centering         
  \includegraphics[width=16cm]{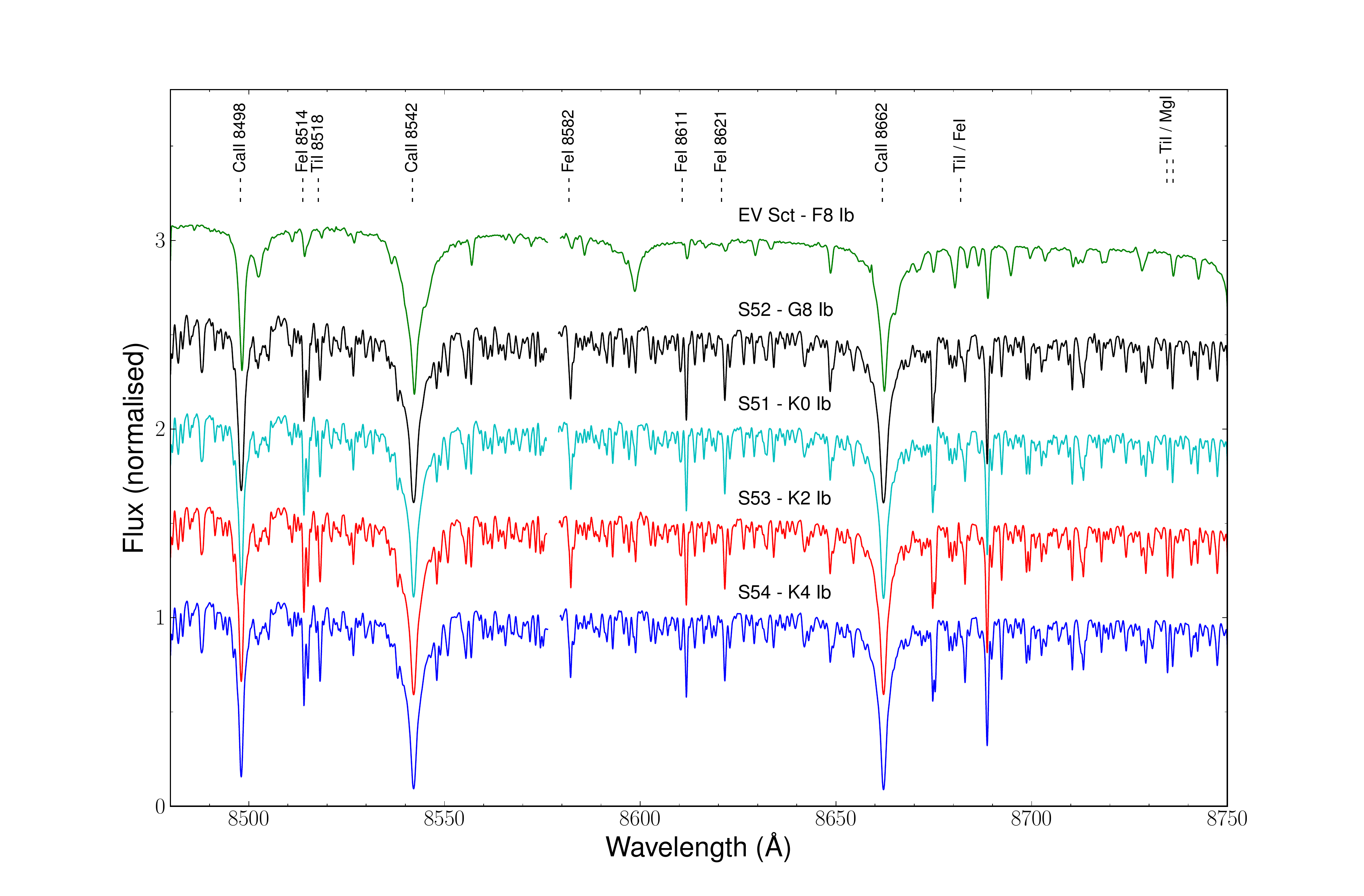}   
  \caption{HERMES spectra for cool stars in the field of NGC\,6664 (Note the gap between orders 41 and 42 around 8580\,\AA).} 
  \label{Rojas_6664}  
\end{figure*}

\textit{NGC\,6664:} We observed 16 stars in the cluster field. Most of them, 11, are B stars ranging types from B2 to B9. The five remaining stars are evolved objects: 
the Cepheid EV\,Sct and four GK red (super)giants, namely S51, S52, S53 and S54. Figure\,\ref{Rojas_6664} displays their spectra. 
Among the B stars, we found two Be stars (S56 and S228). \citet{McSw09}, found five Be stars in this cluster from spectroscopic observations focused around the H$\alpha$ line.
Two of them, S56 and S61, were also observed by us. The first star, S56, exhibits emission in our spectrum while the other one, S61 does not. It is important to remember that 
our spectra do not cover H$\alpha$ but H$\beta$. S61 might still show emission only around H$\alpha$. Nevertheless, in a previous study \citet{McSw05}, by using H$\alpha$ photometry,
also found no emission in this star. The other Be star identified in this work, S228, was already included in the catalogue of Be stars from \citet{SS77} with the number 401.

\textit{Be\,55:} In this case we observed 12 stars out of which five are B-type stars. These stars are among the earliest in the cluster, with spectral types in the range B4-B6.
None of them exhibit an emission profile in the Balmer lines. The remaining seven stars are cool objects with spectral types comprised between G8 and M2, all 
of them classified as supergiants.

\subsection{Cluster membership}\label{membership}

Identifying the members associated with a cluster among all the stars present in the field is essential to characterise the cluster. Astrometry is an efficient 
tool to perform this disentanglement. $Gaia$ is beginning to revolutionise astronomy by providing the most accurate astrometry for the largest number of objects 
to date. Based on the DR2 data, \cite{cantat18} studied a large sample of Galactic open clusters, assigning the membership probabilities for thousands of stars.
In this work we analysed the aforementioned clusters taking into account the members found by \cite{cantat18}, paying special attention to the evolved stars.
The membership of all stars observed in this work are summarised in Table~\ref{tab_mp}.

\textit{NGC\,6649:} \cite{Me08} observed in this cluster five evolved stars: three RSGs (S42, S49 and S117) and two YSGs (S19 and the Cepheid, S64). All of them, 
with the exception of S42, are bona-fide members. This star, as already noted by \cite{Ma75}, is a double object composed of a blue (S42\,B) and a red star (S42\,K).
The latest is the one observed by \cite{Me08}. The membership of S42\,B is also discarded, a fact expected from photometry (see Fig.~\ref{CMD_6649}) since despite having the 
latest spectral type among the blue stars, it is the brightest one. Finally, the astrometry of the coolest target observed, S111, is not compatible with membership 
(as neither is its RV).

\textit{NGC\,6664:} also in this case five evolved stars were observed by \cite{Me08}: the Cepheid EV\,Sct (S80) and four RSGs, namely S51, S52, S53 and S54. According to
\cite{cantat18} only S51, S52 and S80 are likely members. In addition, the colour-magnitude diagram (CMD) shows a star, BD-08\,4641, which might be a new YSG member (see Sect.~\ref{sec_cmd_6664}).

\textit{Be\,55:} This cluster was not observed by \cite{Me08} but \cite{Be55} found in it seven evolved stars. Six of them, S1, S2, S3, S4, S6 and S61, are RSGs while the remaining one, S5, is a YSG \citep[the new Cepheid discovered by][]{Lohr18}.  
The $Gaia$ DR2 astrometry confirms the membership of all these stars, with the only exception of S61, the coolest object observed in the cluster field.

\subsection{Colour-magnitude diagrams}

In order to determine accurate ages for our clusters we combined archival photometry with the spectroscopy obtained in this work. In a first step, we constructed 
for each cluster CMDs in different photometric systems, highlighting the stars for which we have spectra. 
For both optical and 2MASS photometry we plot the dereddened CMDs, i.e. $M_V/(B-V)_0$ and $M_{K_{\textrm{S}}}/(J-K_{\textrm{S}})_0$, respectively, whereas
for the $Gaia$ DR2 photometry we represented the $G/(G_{BP}-G_{RP})$ CMD reddening, instead, the isochrone.
Then, in a second step, we drew PARSEC isochrones \citep{parsec} computed at the corresponding metallicity found in Sect.\,\ref{sec_param}. With the aim of ensuring the reliability of the fitting
we selected only those stars from \cite{cantat18} with a membership probability sufficiently high (i.e. $P\geq$\,0.7).

\subsubsection{NGC\,6649}

We performed the analysis of the CCD photometry from \citet{Wa87} since that from \citet{Hoy03}, more modern, is not publicly available. It covers a field of $2\farcm7$ 
around the nominal cluster centre, with a magnitude limit $V\approx20$. There is $BV$ photometry for 395 stars, out of which 82 are also observed in the $U$ band. 
This photometry does not provide any data for certain interesting stars such as S52, S111, the Cepheid and the components of the binary S42, the brightest in the cluster field. 
For the last three, we included their values from \citet{Ma75} in our sample after rescaling the photoelectric $UBV$ values taking into account the offset between both photometric datasets. 

\begin{figure}  
  \centering         
  \includegraphics[width=\columnwidth]{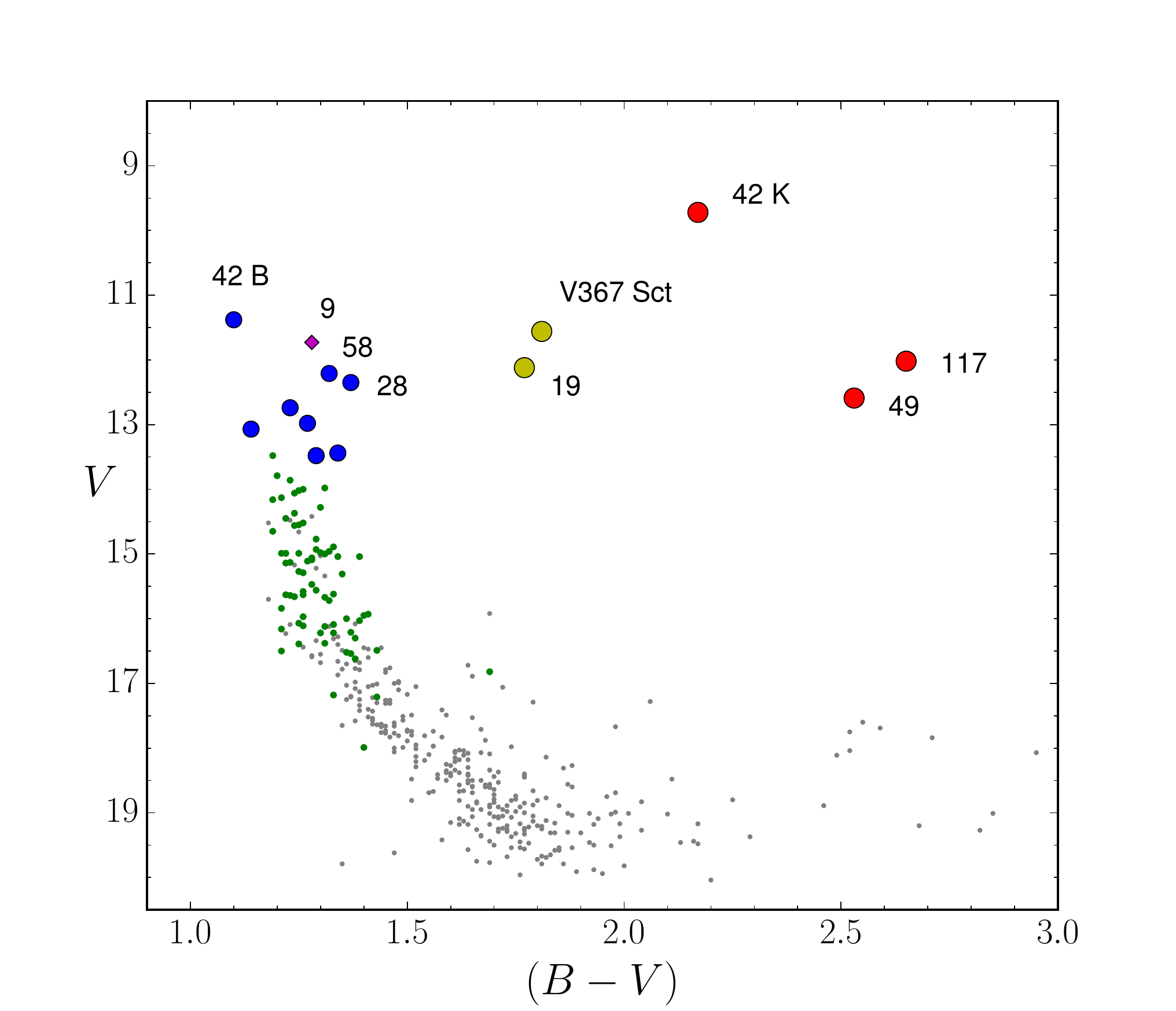}   
  \caption{$V/(B-V)$ diagram for all stars in the field of NGC\,6649. Stars with only $BV$ photometry are represented by gray dots whereas those observed also in the $U$ band appear as green dots.
  Stars for which, in addition, we have spectra are plotted with small blue circles (B-type stars), magenta diamond (Be star) and big circles: greenish-yellow for YSGs and red for RSGs. The numbering for 
  the most representative stars is marked as well.} 
  \label{CMD_6649}  
\end{figure}

In the first place we plot the $V/(B-V)$ diagram showing also the stars observed spectroscopically (Fig.~\ref{CMD_6649}). To deredden the cluster stars we followed 
the procedure employed in \cite{3105}, based on the classical extinction-free $Q$ parameter \citep{JM53}. Since this parameter is defined as $Q=(U-B)-X(B-V)$,
we used only those stars with photometry also in the $U$ band. From six B-stars without any peculiarities (see Table~\ref{red_6649}) we found a mean reddening of
$E(B-V)$=1.43$\pm$0.09, and determined an average ratio $X$=0.79$\pm$0.04, value slightly different to the standard value, $X$=0.72$\pm$0.03 \citep{JM53}. With this
value we computed the $Q$ index for selecting intrinsically blue stars in the field and assigned them photometric spectral types. 
We checked the validity of this spectral classification by comparing, when possible, the photometric types with those directly observed from spectra. We found consistency
between both classifications within one spectral subtype.

\begin{table*}
  \caption{Colour excesses for B-type stars in the field of NGC\,6649. Intrinsic colours are adopted from \citet{Fi70}.\label{red_6649}}
\begin{center}
\begin{tabular}{lcccccccc}
\hline\hline
Star & Sp T & $(B-V)$ & $(U-B)$ & $(B-V)_0$ & $(U-B)_0$ & $E(B-V)$ & $E(U-B)$ & $X$ \\
\hline
14 & B5\,V   & 1.34 & 0.65 & $-$0.16 & $-$0.58 & 1.50 & 1.23 & 0.82 \\
23 & B7\,III & 1.27 & 0.64 & $-$0.12 & $-$0.44 & 1.39 & 1.08 & 0.78 \\
28 & B4\,III & 1.37 & 0.61 & $-$0.18 & $-$0.59 & 1.55 & 1.20 & 0.77 \\
33 & B6\,IV  & 1.23 & 0.53 & $-$0.14 & $-$0.47 & 1.37 & 1.00 & 0.73 \\
35 & B5\,V   & 1.14 & 0.47 & $-$0.16 & $-$0.58 & 1.30 & 1.05 & 0.81 \\
48 & B5\,V   & 1.29 & 0.65 & $-$0.16 & $-$0.58 & 1.45 & 1.23 & 0.85 \\
\hline

\end{tabular}
\end{center}
\end{table*}

\begin{figure*}
  \centering         
  \includegraphics[width=17cm]{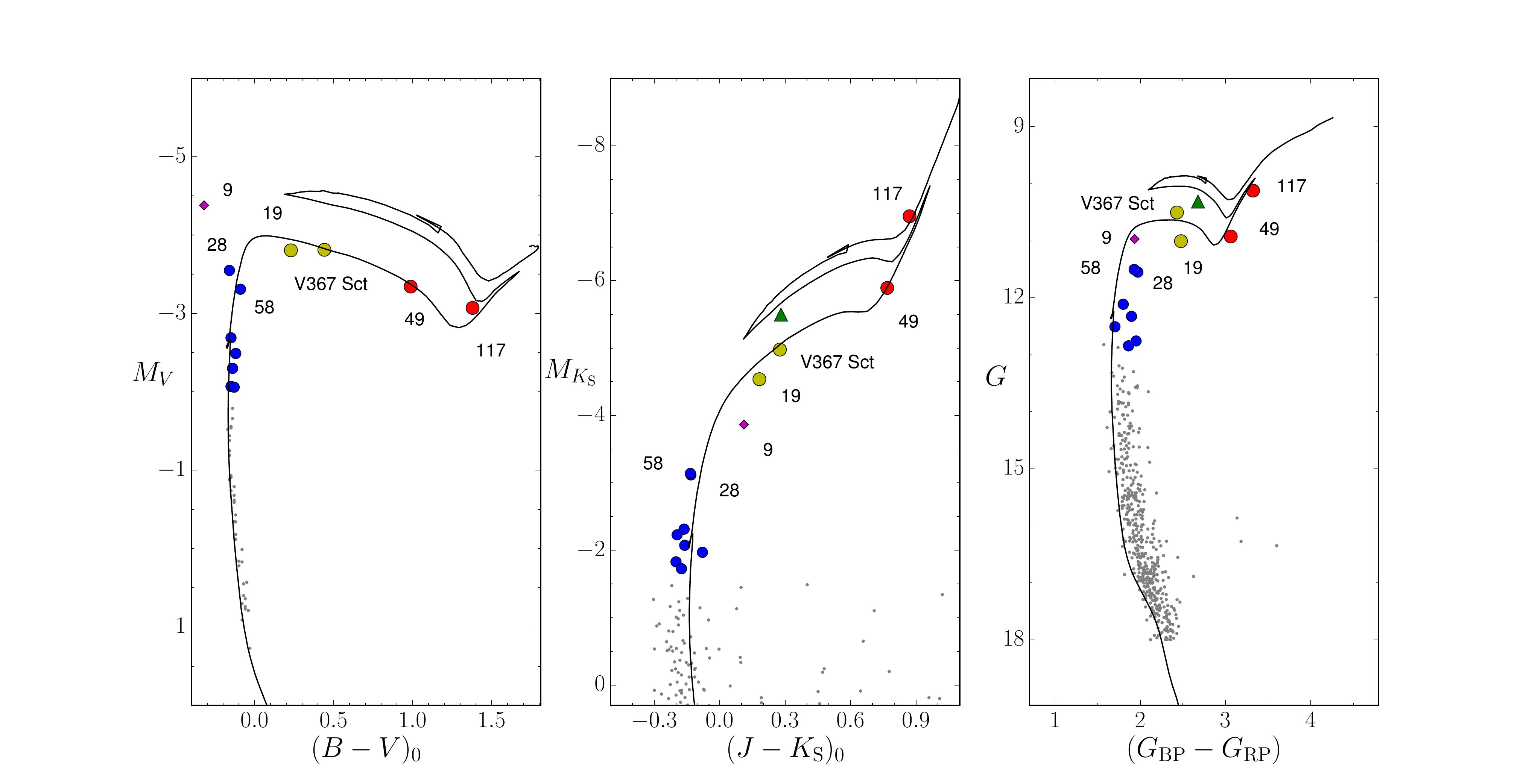}   
  \caption{Colour-magnitude diagrams for likely members of NGC\,6649 in three different photometric systems: \textbf{Left:} $M_V/(B-V)_0$ from the optical photometry taken by \citet{Wa87};  
  \textbf{Centre:} $M_{K_{\textrm{S}}}/(J-K_{\textrm{S}})_0$ (2MASS photometry) and \textbf{Right:} $G/(G_{BP}-G_{RP})$ ($Gaia$ DR2 data). Grey dots are the photometric data for likely members.
  Stars observed spectroscopically are represented as red circles (RSGs), greenish-yellow circles (YSGs), blue circles (blue stars) and magenta diamonds (Be stars). The black line shows the 
  best-fitting PARSEC isochrone (log\,$\tau$=7.8). Star 2MASS J18335423-1019100, a possible new evolved member, is highlighted as a green triangle.}
  \label{isoc_6649}
\end{figure*}

Based on these spectral types and the positions of the stars on the $V/(B-V)$ diagram, we selected the likely cluster members, 44 in total. From the $Q$ index for all 
photometric cluster members we estimate a mean reddening for this cluster of $E(B-V)$=1.39$\pm$0.06, compatible within the errors with that previously estimated from spectra.
For the evolved population, in order to correct them from reddening we followed the procedure described in \citet{fernie63}, more specific for this sort of (super)giants.
Once we had done the member selection, we estimated the cluster distance by a visual ZAMS fitting. The distance modulus obtained is $\mu$=$V_0-M_V$=11.15$\pm$0.15, corresponding 
to a distance of $d$=1.70$\pm$0.12\,kpc. The error involves the uncertainty when considering the ZAMS as a lower envelope. Our estimate, result of the photometric analysis, 
is consistent with that derived (2.0$\pm$0.4\,kpc) from the $Gaia$ DR2 parallax \citep[0.467$\pm$0.087\,mas, according to][where the error represents the dispersion of individual parallaxes among members]{cantat18}.
When converting the parallax into a distance the correction proposed by \citet{lindegren18}, i. e. the addition of 0.029 mas to the published value
with the aim of counteracting the zero-point offset of the $Gaia$ DR2 parallaxes, has been taken into account.
As reddening and distance have been fixed we proceed to determine the cluster age. For this task we carefully fitted by eye different isochrones on the $M_V/(B-V)_0$ diagram. The best-fitting isochrone 
(see Fig.~\ref{isoc_6649}) yields a log\,$\tau$=7.8$\pm$0.1, which is equivalent to an age of 63$\pm$15\,Ma. In this case the error reflects the interval of isochrones which gives a good fit. All
stars fit pretty well the isochrone. Only S9, an extreme Be star, lies away. In addition,  this object is an X-ray source and it is a candidate to be 
a blue straggler star \citep[BSS,][]{amparo_6649}. In addition, the position of S28, already reported as a Be star by \citet{Ma75}, makes it suspected of being another BSS, although 
this location might also be conditionated by its Be nature.
In Fig.~\ref{isoc_6649} we also plot the 2MASS $K_{\textrm {S}}$/$(J-K_{\textrm{S}})_0$ and $Gaia$ DR2 $G/(G_{BP}-G_{RP})$ diagrams, obtaining a result analogous to that derived from the optical CMD.
In these diagrams, the 484 objects found with the highest probability of membership \citep[$P\geq$\,0.7, according to][]{cantat18} have been added, together with their 2MASS counterparts, to the stars observed 
spectroscopically in this work. 
We note the presence of star 2MASS J18335423-1019100 (green triangle in Fig.~\ref{isoc_6649}) in the evolved region on the diagrams, brighter than Cepheid V367\,Sct ($K_{\textrm{s}}=6.2$).
This object is located 8.4$\arcmin$ away from the nominal cluster centre, reason for which it was not covered by the optical photometry.

\subsubsection{NGC\,6664}\label{sec_cmd_6664}

\begin{figure*}
  \centering         
  \includegraphics[width=17cm]{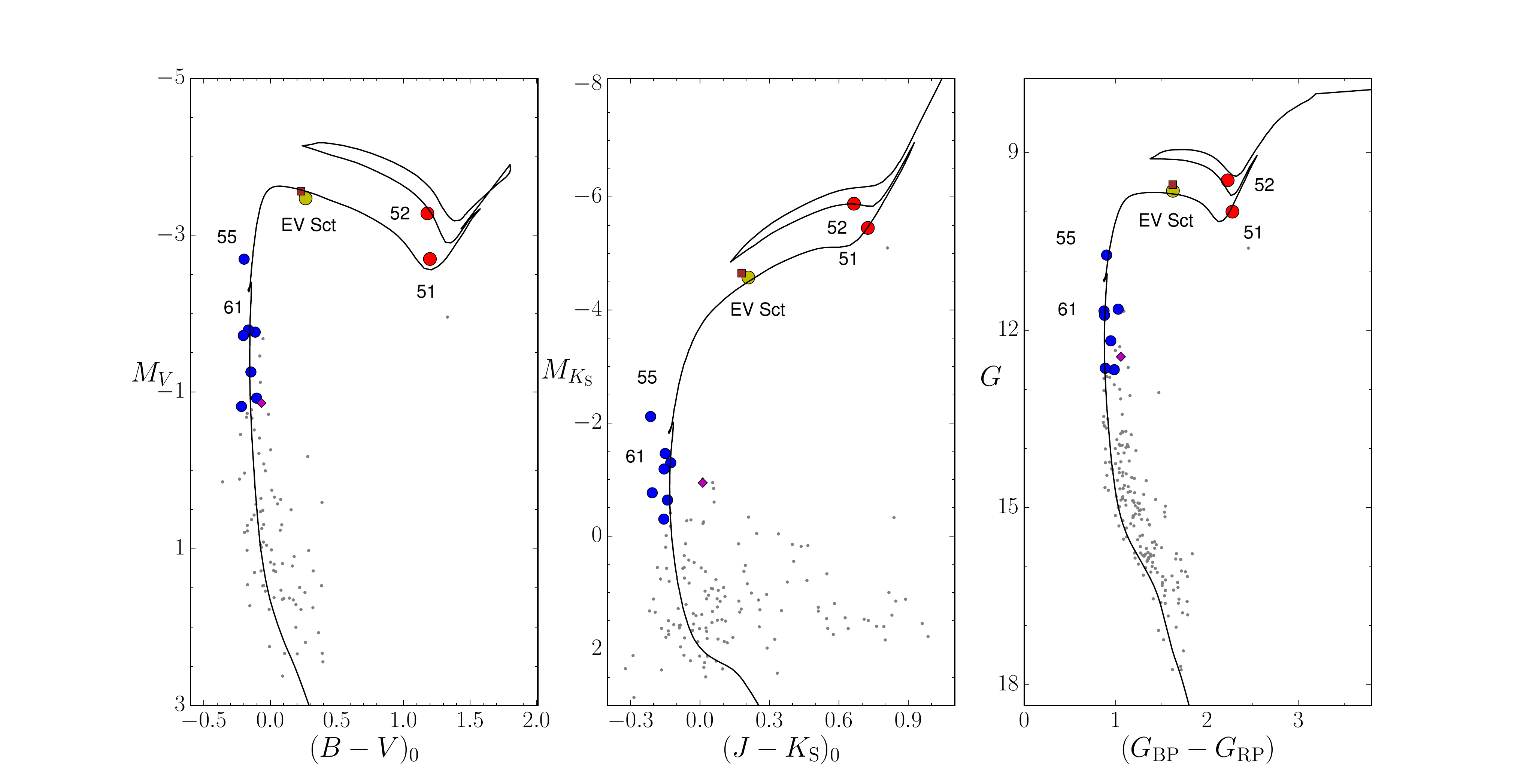}   
  \caption{Colour-magnitude diagrams for likely members of NGC\,6664. Colours and symbols are the same as those in Fig.~\ref{isoc_6649}. The black line shows the best-fitting
  PARSEC isochrone (log\,$\tau$=7.90). Star BD-08 4641, a good candidate to be a new YSG cluster member, is highlighted as a brown square very close to the Cepheid
  EV\,Sct.}
  \label{isoc_6664}
\end{figure*}

In the case of this cluster we resorted to $BV$ photometry from the APASS catalogue \citep[DR9,][]{apass} because that of \citet{Hoy03} is not publicly available.
In order to estimate the cluster reddening we averaged the individual values of the blue stars (without emission lines) observed spectroscopically. By comparing the $(B-V)$ colour for 
each star with the one displayed in the calibration of \citet{Fi70} according to its spectral type, we obtained the individual reddenings (Table\,\ref{red_6664}).
The mean value, $E(B-V)=0.77\pm0.05$, is compatible within the errors with the previous estimates (see Sect.\,\ref{intro_6664}). In addition, this value 
is also compatible with that derived by using the calibration of \citet{St09} from 2MASS photometry. In this case we obtained $E(J-K_{\textrm{s}})=0.36\pm0.03$, which is 
equivalent to $E(B-V)\approx0.72$. Taking into consideration this value we plot the CMDs from the stars observed by us together with the 180 high-probability members (with their APASS
and 2MASS counterparts) found in the list of \citet{cantat18}. The best fit, displayed in Fig.~\ref{isoc_6664},
corresponds to a distance modulus, $\mu$=11.25$\pm$0.15 (i.e. 1.78$\pm$0.12\,kpc), compatible within the errors with the value derived by \citet[][$d$=2.0$\pm$0.2\,kpc]{cantat18}, and 
an age, log\,$\tau$=7.90$\pm$0.10 (i.e. 79$\pm$18\,Ma). In this diagram all likely members lie on the isochrone.
However, it is worth noting the spread around one magnitude observed among the blue stars at the top of the MS. This fact could indicate the existence of a 
variable reddening across the field. 

As in the previous cluster, the presence of three BSS candidates in the field of NGC\,6664 is reported in the literature \citep{scm6664,Ah95}. Only two of them, S55 and S61, 
are likely members and their positions are displayed on the diagrams. This issue will be discussed in more detail later in Sect.\,\ref{sec_BSS}.
Additionally, we note that the position on the CMDs of 
two objects suggests that these might be new evolved members. The first one is the star BD-08\,4641. Since it is located at almost the same position as the Cepheid it would be 
a new likely YSG member. The second candidate is the object TYC\,5691-1067-1. On the CMDs it occupies a location below S51 (see Fig.~\ref{isoc_6664}),
in the reddest part of the diagrams. However, it is not as close to the isochrone as would be expected from a bona-fide member. In addition, both stars are placed away 
from the cluster centre (16.2\arcmin and 20.2\arcmin, respectively).

\begin{table}
  \caption{Colour excesses for B-type stars in the field of NGC\,6664. Intrinsic colours are adopted from \citet{Fi70}.\label{red_6664}}
\begin{center}
\begin{tabular}{lcccccccc}
\hline\hline
Star & Sp T & $V$ & $(B-V)$ & $(B-V)_0$ & $E(B-V)$ \\
\hline
55 & BN2\,IV  & 10.94 & 0.57 & $-$0.24 & 0.81 \\
60 & B6\,IV   & 11.87 & 0.66 & $-$0.14 & 0.80 \\
61 & B3\,IV   & 11.85 & 0.61 & $-$0.20 & 0.81 \\
62 & B6\,IV   & 11.92 & 0.57 & $-$0.14 & 0.71 \\
63 & B5\,III  & 12.38 & 0.62 & $-$0.16 & 0.78 \\
64 & B6\,V    & 12.72 & 0.67 & $-$0.14 & 0.81 \\
A  & B6\,IV   & 12.82 & 0.55 & $-$0.14 & 0.69 \\
\hline

\end{tabular}
\end{center}
\end{table}

\subsubsection{Be\,55}

\begin{figure*}
  \centering         
  \includegraphics[width=17cm]{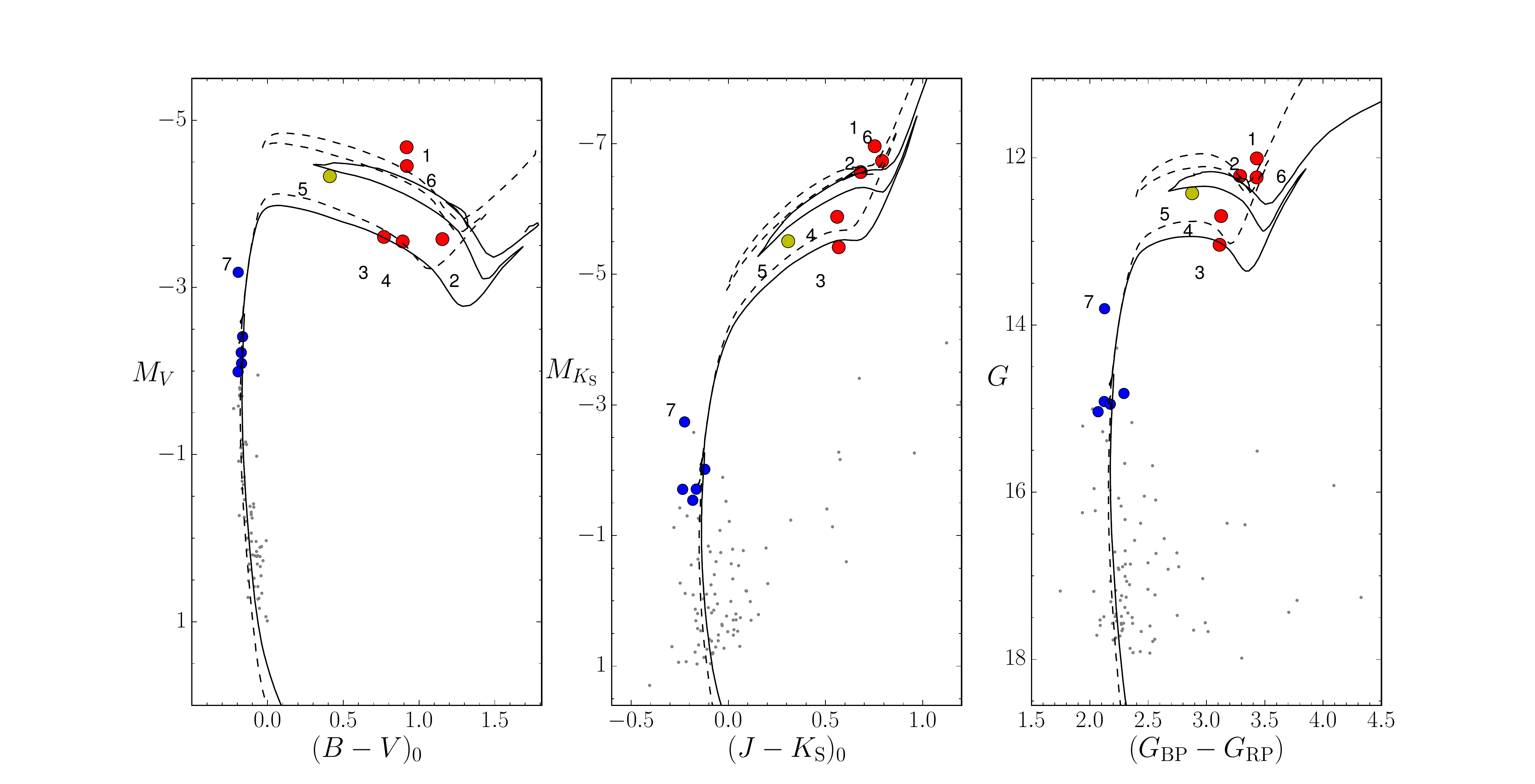}   
  \caption{Colour-magnitude diagrams for likely members of Be\,55. Colours and symbols are the same as those in Fig.~\ref{isoc_6649}. The solid line shows the best-fitting
  PARSEC isochrone (at the Cepheid metallicity) whereas the dashed line corresponds to that computed at [Fe/H]=$-$0.4 (see text for explanation).} 
  \label{isoc_be55}
\end{figure*}

This cluster was already observed by our group \citep{Be55}. From the analysis of $UBV$ photometry we identified 138 likely members for which we computed its reddening. 
Among these, half of them are bona-fide members according to \citet{cantat18}.
By averaging their individual reddenings we obtained the value for the cluster, $E(B-V)$=1.81$\pm$0.15. Given its large dispersion it is reasonable to assume the 
existence of a non-negligible differential reddening in the field. We constructed the CMD taking into account the evolved 
objects too. For one of them, S1, our photometry did not provide any data, so we took the values given by \cite{Maciejewski07} for this star, after correcting 
them from the average differences existing between both photometric datasets. 
In Fig.~\ref{isoc_be55}, from 107 high-probability members, the optical (left), 2MASS (centre) and $Gaia$ DR2 (right) CMDs are displayed.
The best-fitting isochrone, computed at [Fe/H]=+0.07 \citep[the value found for the Cepheid,][]{Lohr18}, corresponds to a log\,$\tau$=7.80$\pm$0.10, which is equivalent to a 
cluster age of 63$\pm$15\,Ma, and a distance, $\mu$=12.55$\pm$0.15 (i.e. 3.24$\pm$0.22\,kpc), consistent with that obtained by \citet[][i.e. 3.0$\pm$0.8\,kpc]{cantat18} from an 
average cluster parallax of 0.309$\pm$0.091\,mas. 
On the one hand, blue members lie on the isochrone in the three CMDs and, given the position of S7, it appears to be a BSS candidate. On the other hand, evolved stars do not show 
a good fit: their location does not agree with their spectral types, all of them are displaced to the left with respect to the expected position. 
This is surely an effect of the strong reddening present in the field ($A_{V}\approx$5.5 mag.), which makes it difficult to correct properly for the evolved members.
This effect is more important in the optical CMD \citep[despite the correction followed,][]{fernie63} than in the 2MASS CMD, less sensitive to the reddening. Since it is 
also visible in the $Gaia$ DR2 CMD, in which the isochrone (and not the stars) was automatically reddened, it seems that the reddening derived from blue stars is not the problem.
Another possible explanation could be metallicity. It is well known that the lower the cluster metallicity, the warmer its evolved stars.
In order to test this possibility we looked for isochrones with different metallicities, finding that a value of [Fe/H]$\approx-$0.4 yields the one that matches the position
of the evolved stars better (see Fig.~\ref{isoc_be55}). However, according to its Galactocentric position ($R_{\textrm{GC}}\approx$8.5\,kpc), this value does not seem to be very reliable 
and, therefore, the reddening appears to be the most probable cause.

\subsection{Size and mass of the clusters}\label{sec_size_mass}

With the aim of determining the size of the clusters under study we resorted to $Gaia$ DR2 data. Since it is an all-sky survey, it allows us to inspect not only the 
central part of the cluster but also its surroundings, which gives us the possibility to investigate its extent. We started our analysis selecting the $Gaia$ DR2 sources 
inside a radius of $30\arcmin$ around the nominal cluster centre. Due to the huge number of objects retrieved, we modified our selection to keep only those objects 
with better astrometry (i. e. a parallax error smaller than 0.15 mas). In a first step, we calculated the position of the cluster centre. We assumed that it is the region 
where the stellar density is higher. In this way, we found it as the peak of the Gaussian corresponding to the fitting of the density profile observed along each equatorial 
coordinate. In a second step, we determined the stellar projected distribution by counting stars in concentric annuli around the new centre. Then, we fitted this density 
profile to a classical three-parameter King-model \citep{king}, obtaining the cluster size in terms of the core and tidal radii. The first one is defined as the radial distance
at which the density becomes half of the central value whereas the latter refers to the distance at which the cluster is diluted in the stellar background.

\begin{figure*}
  \centering         
  \includegraphics[width=19cm]{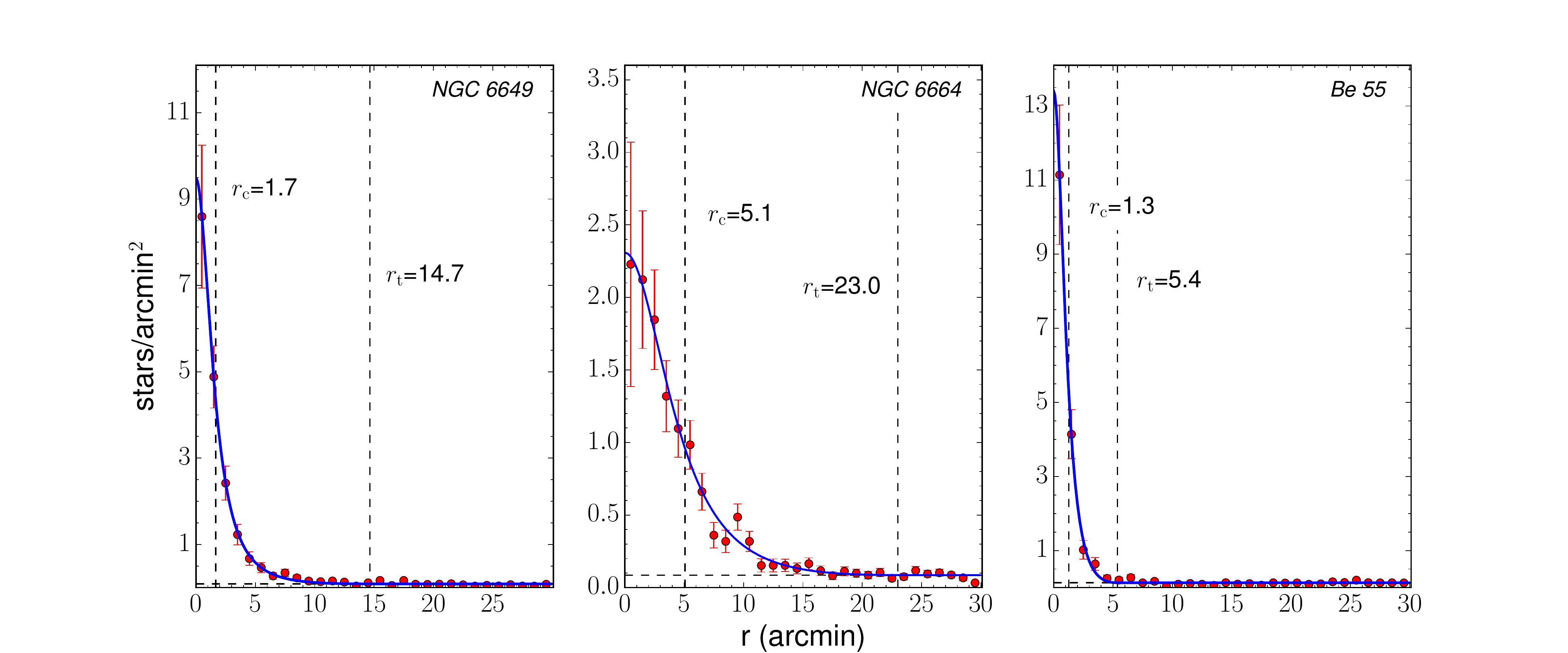}   
  \caption{Projected density distribution around the clusters studied in this work. Red circles are the observed values together with the Poisson errors whereas 
  the blue line shows the fitted King profile. Vertical dashed lines represents the core ($r_{\textrm{c}}$) and tidal ($r_{\textrm{t}}$) radii while the horizontal one is the background 
  density.}
  \label{fig_king}
\end{figure*}

The region in the sky where each cluster is located as well as the properties of the cluster itself determine how it stands out from the field. 
In order to facilitate the identification of the cluster we payed attention to the distribution of the brightest stars in the field (i.e. $G\leq$16), which
are often useful as tracers of the cluster boundary. In conjunction with it, we also examined the arrengement of stars with a similar astrometry to cluster members, i. e., 
those objects that in the astrometric space ($\varpi$, $\mu_{\alpha*}$, $\mu_{\delta}$) are within a 3-$\sigma$ radius around the cluster centre reported by \citet{cantat18}.
Our results, coordinates of the centre and radii (in both angular and physical units) are displayed in Table~\ref{tab_king}. We noted that NGC\,6649 and Be\,55 are compact clusters easily distinguishable from the field. 
Significant differences with the nominal centres are not found (this work$-$nominal), $\Delta(\alpha,\delta)$=(1.7$^s$,$-$12.2$^{\arcsec}$) for NGC\,6649 and 
$\Delta(\alpha,\delta)$=(2.4$^s$,$-$8.6$^{\arcsec}$) for Be\,55. However, in the case of NGC\,6664 the opposite is true. The cluster does not stand out from the environment and, 
therefore, its characterisation becomes more complicated, which leads to larger uncertainties. In this case, it is appreciated a difference with respect to the nominal 
centre of $\Delta(\alpha,\delta)$=(11.3$^s$,$-$80.2$^{\arcsec}$), considerably further than the previous values. Figure~\ref{fig_king} shows the King profiles 
fitted for each cluster indicating the position of the core and tidal radii. We will discuss about it in Sect.~\ref{sec_clust_param}.

\begin{table*}
  \caption{Equatorial coordinates (J2000) for the centre and core ($r_{\textrm{c}}$) and tidal ($r_{\textrm{t}}$) radii for the studied clusters.\label{tab_king}}
\begin{center}
\begin{tabular}{l|cc|cc|cc}
\hline\hline
Cluster   & RA & DEC  & $r_{\textrm{c}}$ ($\arcmin$) & $r_{\textrm{t}}$ ($\arcmin$) & $r_{\textrm{c}}$ (pc) & $r_{\textrm{t}}$ (pc) \\
\hline
NGC\,6649 & 278.3554$\pm$0.0029 & $-$10.3999$\pm$0.0022 & 1.67$\pm$0.03 & 14.7$\pm$1.4 & 1.0$\pm$0.2 &  8.6$\pm$2.5 \\
NGC\,6664 & 279.1278$\pm$0.0054 &  $-$8.1944$\pm$0.0071 & 5.08$\pm$0.41 & 23.0$\pm$5.0 & 3.0$\pm$0.7 & 13.7$\pm$4.8 \\
Be\,55    & 319.2319$\pm$0.0017 &    51.7613$\pm$0.0016 & 1.28$\pm$0.06 &  5.4$\pm$0.4 & 1.1$\pm$0.4 &  4.6$\pm$1.9 \\
\hline

\end{tabular}
\end{center}
\end{table*}

Once the size of the cluster is set, its mass can be inferred by integrating the initial mass function \citep[IMF,][]{Kroupa}. This procedure was 
followed in the past in the analysis of other clusters \citep[a more detailed explanation is presented in][]{3105}. With the aim of calibrating the IMF we need to 
know the number of stars, inside the cluster limits, within a certain mass range. In this range completeness must be assured. The best approach to do it is by counting the brightest 
blue stars located at the top of the MS. Our selection spans the whole B spectral type starting from the earliest type observed at 
the turn-off point (TO), B5--B6 in these clusters. From the $Gaia$ DR2 CMD we determined the average $G$ magnitude for these stars ($G_{\textrm{TO}}$) together with the individual dispersion around it. 
Based on the differences in magnitude and mass among the B spectral types \citep[according to][]{st92}, and considering the similarity between the $G$ and $V$ bands, we obtained 
the magnitude range to cover stars from the TO down to spectral type B9\,V (whose average mass is 2.6\,M$_{\sun}$). Then, once the calibration of the IMF was done, its integration
allowed us to obtain an estimate of the cluster mass. Taking into account the assumptions made, this is only an approximate value. These are the numbers for each cluster:

\textit{NGC\,6649}: We find a $G_{\textrm{TO}}$=12.6$\pm$0.2 for a spectral type B5\,V (4.8\,M$_{\sun}$). We cut at $G$=14.1 to reach B9 stars. In total, in 
this range we counted 57 stars which implies a present mass around 1\,800\,M$_{\sun}$, equivalent to an initial mass $\approx$2\,600\,M$_{\sun}$.

\textit{NGC\,6664}: In this case we find a TO around B5--B6\,V, with an average $G_{\textrm{TO}}$=12.2$\pm$0.5. From 61 stars (down to $G$=13.9) we inferred a present mass around 2\,000\,M$_{\sun}$ and
an initial value 2\,900M\,$_{\sun}$. 

\textit{Be\,55}: This is the least massive cluster. Only 24 stars are observed between B6\,V (at $G_{\textrm{TO}}$=14.9$\pm$0.1 and 4.1\,M$_{\sun}$) and B9\,V ($G$=16.0). This number
implies a mass of 900\,M\,$_{\sun}$ for the present and $\approx$1\,300\,M$_{\sun}$ for its birth. We will come back to this in Sect.~\ref{sec_clust_param}.

\subsection{Spectroscopic analysis}

\subsubsection{Radial and rotational velocities}\label{sec_vrad}

For the cool stars (observed at high resolution), before computing the atmospheric parameters and the chemical abundances, we need to prepare their spectra.
In a first step, we corrected them for the RV. We determined the heliocentric radial velocity through Fourier cross-correlation by employing the {\scshape iSpec} software \citep{ispec}.
For each target spectrum the cross-correlation was computed against a list of atomic lines masks selected for the {\em Gaia} benchmark stars library pipeline, from asteroids observed with the
NARVAL spectrograph.

\textit{NGC\,6649:} For this cluster we obtained the mean RV from the two red (super)giants (i.e. S49 and S117) since the RV of the Cepheid is somewhat different from those stars 
(probably due to its variability, as this number, derived from a single epoch, only represents a random point of its radial velocity curve). Our value, $v_{\textrm{rad}}$\,=$-$8.66\,$\pm$\,1.03\,\,km\,s$^{-1}$, is in perfect agreement with 8.59\,$\pm$\,0.40\,km\,s$^{-1}$, the value computed by \citet{Me08}
from five stars (including the Cepheid and the non member S42\,K). The non membership of S42\,K and S111 (see Table~\ref{par_cool}) is confirmed also based on a 
RV criterion. 

\textit{NGC\,6664:} EV\,Sct shows a noisy double-peak CCF, despite not being a binary object (as noted above in Sect.\,~\ref{intro_6664}). This fact prevents us from carrying out a
reliable analysis of its spectrum, reason for which this star will not be taken into account to estimate the cluster average.
Among the remaining evolved stars, S51 and S52, we calculated the RV for the cluster, $v_{\textrm{rad}}$\,=19.34\,$\pm$\,0.21\,\,km\,s$^{-1}$. This value is compatible, within the errors, 
with the value obtained by \citet[][, 18.58\,$\pm$\,0.69\,\,km\,s$^{-1}$]{Me08}.

\textit{Be\,55:} The resolution at which we observed the stars of this cluster does not allow us to perform the spectroscopic analysis, but it is enough to estimate their RVs from the 
IDS spectra (those with the highest resolution available in our sample, displayed in Table~\ref{vrad_be55}). We obtained a (weighted) average value for the cluster, $v_{\textrm{rad}}$\,=$-$31.7\,$\pm$\,7.4\,\,km\,s$^{-1}$. 
As shown, the error (calculated as the dispersion of individual measurements) is quite large, since RVs for bona-fide members are distributed in a wide range ($\approx$\,$-$20--40\,km\,s$^{-1}$). 
Star S1 is responsible for much of this dispersion and, additionally, it is the brightest star of the cluster. This fact could be indicating its possible binary nature.
Regardless of this star, the average value is $v_{\textrm{rad}}$\,=$-$27.7\,$\pm$\,4.9\,\,km\,s$^{-1}$, whose dispersion is within twice the instrumental
error (around 3\,km\,s$^{-1}$).
Unfortunately, Be\,55, unlike the previous clusters, was not observed by \citet{Me08}. 

\begin{table}
\caption{Radial velocites for stars in the field of Be\,55.\label{vrad_be55}}
\begin{center}
\begin{tabular}{lcc}   
\hline\hline
Star & Sp T & RV (km\,s$^{-1}$) \\
\hline
1          & K1\,Ib    &  $-$42.61 $\pm$ 0.37 \\ 
2          & K0\,Ib    &  $-$24.55 $\pm$ 0.39 \\ 
3          & G8\,II    &  $-$33.33 $\pm$ 0.47 \\ 
4          & K0\,Ib-II &  $-$28.97 $\pm$ 0.47 \\ 
5          & F8\,Ib    &  $-$31.63 $\pm$ 0.86 \\ 
6          & K4\,II    &  $-$21.61 $\pm$ 0.60 \\ 
\textit{61}$^{*}$   & M2\,II    &  $-$30.30 $\pm$ 0.85 \\ 
\hline

\end{tabular}
\end{center}
\end{table}

In a second step we estimated the contribution of rotational and macroturbulence broadenings since the former is used as an input when computing the atmospheric parameters. 
The projected rotational velocity ($v \sin\,i$) is calculated by using the {\scshape iacob-broad} code \citep{iacob_broad}. It is based on the Fourier transform method and is able to separate rotation
from other broadening mechanisms such as the macroturbulent velocity ($\zeta$). We used as a diagnostic, at least, eight lines of \ion{Fe}{i} and \ion{Ni}{i} clearly visible in the spectra.
The errors reflect the scatter between measurements, in terms of rms. All these radial and rotational velocities, for NGC\,6649 and NGC\,6664 are listed in Table~\ref{par_cool}.

\subsubsection{Stellar atmospheric parameters}\label{sec_param}

We derived the stellar atmospheric parameters for 35 objects. However, as we have done in previous works \citep[see e. g.][]{2345}, depending on the 
temperature range of the stars two different procedures are followed. Additionally, it is important to note the different resolution with which both 
groups of targets have been observed. On the one hand we have low-resolution spectra for B stars whereas on the other hand, high-resolution spectra have
been taken for the cool stars.

Regarding B stars, we computed stellar parameters from their ISIS spectra for 26 stars distributed 10 in the field of NGC\,6649, 11 in the one of 
NGC\,6664 and the remaining five in Be\,55. We followed the strategy described by \citet{Ca12}, by employing a grid of {\scshape fastwind} synthetic spectra
\citep{Si11,Ca12} and adopting {\scshape fastwind} as the stellar atmosphere code \citep{Sa97,Pu05}. 
By using an automatic $\chi^2$-based algorithm we found the stellar atmospheric parameters  that best reproduce the main strong features observed in the range $\approx$\,4000\,--\,5000\,\AA{}. 
Because of the resolution of these spectra is not as high as that necessary to separate the different broadenings, we assumed a rotational origin for the whole
broadening. Hence, we also calculated the rotational velocity in an interactive way. First, we chose as an initial estimate of the $v\sin\,i$ a value close to the 
resolution, i. e. around 50\,km\,s$^{-1}$. Then, we computed a first model capable of reproducing the spectrum. Once the stellar parameters are fixed, we looked 
for a second value of rotation by changing it until finding a new model which reproduced best the profiles. We repeated this process at least a couple of times, 
to make sure that the rotation does not change. Results (i.e. effective temperature, surface gravity and projected rotational velocity) are displayed in Table~\ref{par_blue}. 
The temperatures derived are in good agreement with the spectral types assigned according to the calibration by \citet{Hump84}.

\begin{table}
\caption{Stellar atmospheric parameters for the blue stars derived from ISIS spectra. Non members are marked with an asterisk.\label{par_blue}}
\begin{center}
\begin{tabular}{lcccc}   
\hline\hline
\multirow{2}{*}{Star} & \multirow{2}{*}{Sp T} & $v \sin\,i$    & $T_{\textrm{eff}}$  & $\log\,g$ \\
                     &                      & (km\,s$^{-1}$) &    (kK)              &   (dex)   \\
\hline
\multicolumn{5}{c}{NGC\,6649}\\
\hline
9    & B1\,IIIe & 170 & 25.0 $\pm$ 2.8 & 3.1 $\pm$ 0.2 \\
14   & B5\,V    & 170 & 13.0 $\pm$ 1.0 & 3.0 $\pm$ 0.2 \\
23   & B7\,III  & 210 & 13.0 $\pm$ 1.0 & 3.1 $\pm$ 0.2 \\
28   & B4\,III  &  50 & 13.0 $\pm$ 1.0 & 2.8 $\pm$ 0.1 \\
33   & B6\,IV   &  50 & 13.0 $\pm$ 1.6 & 3.1 $\pm$ 0.2 \\
35   & B5\,V    & 130 & 14.0 $\pm$ 1.0 & 3.5 $\pm$ 0.1 \\
\textit{42\,B}$^{*,**}$ & B8\,IV & 310 & 13.0 $\pm$ 1.0 & 3.5 $\pm$ 0.2 \\
48   & B5\,V    &  50 & 14.0 $\pm$ 1.0 & 3.6 $\pm$ 0.2 \\
52   & B5\,V    & 250 & 14.0 $\pm$ 1.3 & 3.5 $\pm$ 0.2 \\
58   & B8\,II & 150 & 12.0 $\pm$ 1.0 & 3.2 $\pm$ 0.1 \\
\hline
\multicolumn{5}{c}{NGC\,6664}\\
\hline
\textit{50}$^{*}$  & B9\,IV   & 270  & 12.0 $\pm$ 1.0 & 4.1 $\pm$ 0.1 \\
55  & BN2\,IV  &  50  & 21.0 $\pm$ 1.1 & 3.7 $\pm$ 0.1 \\
\textit{56}$^{*}$  & B6\,IIIe &  50  & 14.0 $\pm$ 1.0 & 3.6 $\pm$ 0.1 \\
\textit{59}$^{*}$  & B2.5\,V  &  70  & 16.0 $\pm$ 1.0 & 3.8 $\pm$ 0.1 \\
60  & B6\,IV   &  50  & 13.0 $\pm$ 1.0 & 3.6 $\pm$ 0.1 \\
61  & B3\,IV   &  50  & 20.0 $\pm$ 1.0 & 3.9 $\pm$ 0.1 \\
62  & B6\,IV   &  50  & 13.0 $\pm$ 1.0 & 3.7 $\pm$ 0.1 \\
63  & B5\,III  &  90  & 14.0 $\pm$ 1.0 & 3.6 $\pm$ 0.1 \\
64  & B6\,V    & 210  & 13.0 $\pm$ 1.0 & 3.7 $\pm$ 0.1 \\
228 & B5\,Ve   & 290  & 15.0 $\pm$ 1.0 & 3.7 $\pm$ 0.1 \\
A   & B6\,IV   & 170  & 13.0 $\pm$ 1.0 & 3.7 $\pm$ 0.1 \\
\hline
\multicolumn{5}{c}{Berkeley\,55}\\
\hline
7   & B4\,IV  &  70 & 16.0 $\pm$ 1.0 & 3.3 $\pm$ 0.2 \\   
10  & B6\,IV  & 110 & 15.0 $\pm$ 1.1 & 3.6 $\pm$ 0.1 \\  
11  & B6\,IV  & 110 & 14.0 $\pm$ 1.0 & 3.5 $\pm$ 0.2 \\  
12  & B6\,IV  & 190 & 15.0 $\pm$ 1.0 & 3.3 $\pm$ 0.1 \\  
17  & B5\,V   &  70 & 16.0 $\pm$ 1.0 & 3.7 $\pm$ 0.1 \\  
\hline

\end{tabular}
\end{center}
\begin{list}{}{}
\item[]$^{**}$ Star observed with FEROS.
  \end{list}
\end{table}

For the cool stars the methodology is quite different. We analysed nine stars, five observed with FEROS in the field of NGC\,6649 and four others in NGC\,6664 observed
with HERMES. Additionally, we also tried to calculate stellar parameters for EV\,Sct. However, because of its double-peak CCF, we have not been able to provide 
reliable results.
We generated a grid of synthetic spectra from LTE MARCS spherical models \citep{marcs} by using {\scshape spectrum} \citep{graco94} as a radiative transfer code.
Effective temperature (T$_{\textrm{eff}}$) ranges from 3\,300~K to 7\,500~K with a step of 100~K up to 4\,000~K and 250~K until 7\,500~K, whereas surface gravity ($\log\,g$)
varies from $-0.5$ to 3.5~dex in 0.5~dex steps. In the case of metallicity (using [Fe/H] as a proxy), the grid covers from $-1.5$ to $+1.0$~dex in~0.25 dex steps.
We fixed the value of the microturbulent velocity ($\xi$) according to the calibration presented in \citet{dutrafe16}. 
Our linelist, based on that provided by \citet{Ge13}, contains $\sim$230 features for \ion{Fe}{i} and $\sim$55 for \ion{Fe}{ii}. 

We employed an updated version of the {\scshape stepar} code \citep{hugo18} tailored to the present problem. 
Our code used the Metropolis--Hastings algorithm as optimization method. It generates simultaneously 48 Markov-chains of 750 points each performing 
a Bayesian parameter estimation. For this task we resorted to an implementation of Goodman $\&$ Weare's Affine Invariant Markov chain Monte Carlo Ensemble sampler \citep{mcmc}.
A $\chi^2$ function is implemented with the aim of fitting the selected iron lines. The stellar rotation and the instrumental broadening are fixed to the value previously derived 
for the former (Sect.~\ref{sec_vrad}) and the resolution of the spectrographs, respectively.
The macroturbulent broadening was left as a free parameter in order to absorb any residual broadening. 

For the M supergiant in NGC\,6649, S111, we were forced to change our methodology. It is known that the spectrum of M stars is dominated by molecular bands that erode the continuum and,
therefore, identify a large number of spectral features becomes very hard, if not impossible.
However, these bands are very useful since their depth is very sensitive to temperature \citep{anib07}. According to this paper, we payed attention to the region 6670--6730\,\AA{},
where the TiO bands at 6681\,\AA{} and 6714\,\AA{} are clearly present. For this star, we computed its stellar parameters based on a $\chi^2$-minimisation code,
with the same grid of MARCS synthentic spectra, previously described, but using {\scshape turbospectrum} \citep{turbo} as a transfer code. 

Results (i.e effective temperature, surface gravity, macroturbulent velocity and iron abundance) are displayed in Table~\ref{par_cool}.
From the analysis of the confirmed evolved members we estimated a solar average metallicity for these clusters. In the case of NGC\,6649, from stars S49, S117 and V367\,Sct,
we computed a [Fe/H]$=+0.02\pm0.07$, calculating it as a weighted average and taking as error the dispersion between individual values. In the same way, 
for NGC\,6664, from S51 and S52, we derived a metallicity, [Fe/H]$=-0.04\pm0.10$.

\begin{table*}
\caption{Stellar atmospheric parameters for the cool stars derived from high-resolution spectra. Next to the cluster name, in brackets the spectrograph used is indicated.
Non members are marked with an asterisk.\label{par_cool}}
\begin{center}
\begin{tabular}{lccccccc}   
\hline\hline
Star & Sp T & $v_{\textrm{rad}}$ (km\,s$^{-1}$) & $v \sin\,i$ (km\,s$^{-1}$) &  $\zeta$ (km\,s$^{-1}$) & $T_{\textrm{eff}}$ (K) & $\log\,g$ & [Fe/H]\\
\hline
\multicolumn{8}{c}{NGC\,6649 (FEROS)}\\
\hline
\textit{42\,K}$^{*}$ & K0\,Ib-II & $-$12.88 $\pm$ 0.02 & 5.8 $\pm$ 0.5 &  4.92 $\pm$ 0.11 & 4\,474 $\pm$ 28  & 1.59 $\pm$ 0.09 & $-$0.04 $\pm$ 0.04 \\
49                   & K1\,Ib    &  $-$9.38 $\pm$ 0.02 & 5.2 $\pm$ 1.0 &  4.88 $\pm$ 0.21 & 4\,181 $\pm$ 48  & 0.98 $\pm$ 0.16 &    0.00 $\pm$ 0.08 \\  
\textit{111}$^{*}$   & M5\,Ib    & $-$40.78 $\pm$ 0.14 &    ---        &  ---             & 3\,800 $\pm$ 100 & 1.0 $\pm$ 0.5   & $-$0.01 $\pm$ 0.25 \\
117                  & K5\,Ib    &  $-$7.93 $\pm$ 0.02 & 6.4 $\pm$ 0.9 &  4.79 $\pm$ 0.23 & 4\,053 $\pm$ 48  & 1.00 $\pm$ 0.14 &    0.08 $\pm$ 0.07 \\
V367 Sct             & F7\,Ib    & $-$22.21 $\pm$ 0.10 & 8.2 $\pm$ 2.1 & 15.06 $\pm$ 0.24 & 5\,875 $\pm$ 57  & 1.80 $\pm$ 0.11 &    0.00 $\pm$ 0.04 \\
\hline
\multicolumn{8}{c}{NGC\,6664 (HERMES)}\\
\hline
51                   & K0\,Ib    &    19.19 $\pm$ 0.02 & 5.6 $\pm$ 1.4 &  7.43 $\pm$ 0.16 & 4\,398 $\pm$ 84  & 1.25 $\pm$ 0.20 & $-$0.10 $\pm$ 0.10 \\
52                   & G8\,Ib    &    19.48 $\pm$ 0.02 & 6.1 $\pm$ 0.8 &  6.24 $\pm$ 0.22 & 4\,208 $\pm$ 60  & 0.92 $\pm$ 0.17 &    0.00 $\pm$ 0.09 \\
\textit{53}$^{*}$    & K2\,Ib    &    13.93 $\pm$ 0.02 & 4.3 $\pm$ 0.5 &  8.98 $\pm$ 0.14 & 3\,960 $\pm$ 42  & 0.68 $\pm$ 0.13 & $-$0.05 $\pm$ 0.07 \\
\textit{54}$^{*}$    & K4\,Ib    &     3.82 $\pm$ 0.02 & 5.5 $\pm$ 1.0 &  8.09 $\pm$ 0.16 & 4\,492 $\pm$ 46  & 1.47 $\pm$ 0.15 &    0.15 $\pm$ 0.07 \\
\hline

\end{tabular}
\end{center}

\end{table*}

\subsubsection{Chemical abundances}

We only derived chemical abundances for the cool stars, eight in total (four in the field of NGC\,6649 and four others in NGC\,6664), since a high spectral resolution
is required for this sort of study. 
Once we set the atmospheric parameters, it is almost trivial to derive the chemical abundances from equivalent widths ($EW$s). For most of the elements analysed, namely, Na, Mg, Si, 
Ca, Ti, Ni, Y, and Ba, we measured the $EW$s using {\scshape tame} \citep{kan12} in a semi-automatic fashion. Instead, for lithium and oxygen, which are more delicate elements,
we measured the $EW$s by hand with the {\scshape iraf} {\scshape splot} task. For the former, we employed a classical analysis using the 6707.8\,\AA{} line, taking into account 
the nearby \ion{Fe}{i} line at 6707.4\,\AA{}. We expressed this abundance in terms of the standard notation, i.e. $A($Li$)= \log\,$[n(Li)/n(H)]\,+\,12.
In the case of oxygen, we followed the procedure described in \citet{bel15} focusing on the [\ion{O}{i}] 6300\,\AA{} line, keeping in mind that it is blended with a \ion{Ni}{i}
feature. Finally, based on the methodology explained in \citet{dor13}, we computed rubidium abundances performing stellar synthesis for the 7800\,\AA{} \ion{Rb}{i} line. 

Results are displayed in Table~\ref{Abund_6649}, for the stars in the field of NGC\,6649, and Table~\ref{Abund_6664}, for NGC\,6664.
We estimate the cluster average by using the weighted arithmetic mean (employing the variances
as weights) taking into account only the members. The error averages the typical individual error and the star-to-star dispersion. 
In both clusters, members share a similar chemical composition. However, in NGC\,6649, V367\,Sct exhibits a Na abundance around 0.7\,dex higher than the other cluster giants. 
Also S117 shows high abundances of Ca and Ni. On the other hand, the largest differences found for members of NGC\,6664 (for O and Si) are around 0.4\,dex.

\begin{table*}
\caption{Chemical abundances, relative to solar abundances by \citet{Gre07}, measured on the cool stars in the field of NGC\,6649.
Non members are marked with an asterisk. }
\begin{center}
\begin{tabular}{lcccccc}   
\hline\hline
Star & [O/H] &[Na/H] & [Mg/H] & [Si/H] & [Ca/H] & [Ti/H]\\
\hline
\textit{42\,K}$^{*}$ &    0.29 $\pm$ 0.07 & $-$0.03 $\pm$ 0.17 &    0.30 $\pm$ 0.03 & $-$0.03 $\pm$ 0.14 & 0.51 $\pm$ 0.08 &    0.44 $\pm$ 0.06 \\
49                   & $-$0.03 $\pm$ 0.11 & $-$0.18 $\pm$ 0.41 &    0.16 $\pm$ 0.05 &    0.43 $\pm$ 0.09 & 0.25 $\pm$ 0.20 &    0.18 $\pm$ 0.17 \\
117                  &    0.09 $\pm$ 0.09 & $-$0.26 $\pm$ 0.04 & $-$0.10 $\pm$ 0.06 &     ---            & 0.62 $\pm$ 0.05 & $-$0.04 $\pm$ 0.16 \\
V367 Sct             &    ---            &    0.51 $\pm$ 0.07 &    0.17 $\pm$ 0.46 &    0.18 $\pm$ 0.12 & 0.06 $\pm$ 0.06 &    0.04 $\pm$ 0.14 \\ 
\hline
Mean          &   0.04 $\pm$ 0.09 & $-$0.10 $\pm$ 0.30  &  0.05 $\pm$ 0.17  &  0.34 $\pm$ 0.14  &  0.40 $\pm$ 0.19  &  0.06 $\pm$ 0.13 \\
              &                   &                     &                   &                   &                   &  \\

\hline\hline
Star & [Ni/H] & [Rb/H] & [Y/H] & [Ba/H] & $EW$(Li) & A(Li)\\
\hline
\textit{42\,K}$^{*}$ & 0.19 $\pm$ 0.14 & $-$0.05         &    0.48 $\pm$ 0.20 & 0.37 $\pm$ 0.08 & 16.0 & $<$\,0.23 \\
49                   & 0.37 $\pm$ 0.11 & $-$0.07         & $-$0.16 $\pm$ 0.42 & 0.52 $\pm$ 0.08 & 42.0 & $<$\,0.19 \\
117                  & 0.68 $\pm$ 0.12 &    0.35         &    0.10 $\pm$ 0.24 & ---             & 63.5 & $<$\,0.27 \\
V367 Sct             &$-$0.06 $\pm$ 0.08 & ---          &    0.19 $\pm$ 0.19 & ---             & ---  & ---       \\ 
\hline
Mean         &  0.22 $\pm$ 0.24   &    0.14 $\pm$ 0.30  &  0.12 $\pm$ 0.23  &  0.52 $\pm$ 0.08  & --- & ---\\
\hline

\end{tabular}

\label{Abund_6649}
\end{center}
\end{table*}

\begin{table*}
\caption{Chemical abundances, relative to solar abundances by \citet{Gre07}, measured on the cool stars in the field of NGC\,6664.
Non members are marked with an asterisk.}
\begin{center}
\begin{tabular}{lcccccc}   
\hline\hline
Star & [O/H] &[Na/H] & [Mg/H] & [Si/H] & [Ca/H] & [Ti/H]\\
\hline
51                &    0.18 $\pm$ 0.15 & $-$0.23 $\pm$ 0.15 & $-$0.01 $\pm$ 0.09 & $-$0.02 $\pm$ 0.15 &    0.04 $\pm$ 0.15 &    0.05 $\pm$ 0.19 \\
52                & $-$0.22 $\pm$ 0.12 & $-$0.15 $\pm$ 0.42 & $-$0.16 $\pm$ 0.06 &    0.47 $\pm$ 0.09 & $-$0.03 $\pm$ 0.19 & $-$0.17 $\pm$ 0.28 \\
\textit{53}$^{*}$ & $-$0.30 $\pm$ 0.04 & $-$0.13 $\pm$ 0.44 &    0.25 $\pm$ 0.01 &    0.47 $\pm$ 0.08 & $-$0.14 $\pm$ 0.09 & $-$0.03 $\pm$ 0.10 \\
\textit{54}$^{*}$ &    0.40 $\pm$ 0.11 & $-$0.07 $\pm$ 0.14 & $-$0.04 $\pm$ 0.06 &    0.27 $\pm$ 0.19 &    0.45 $\pm$ 0.09 &    0.37 $\pm$ 0.16 \\

\hline
Mean & $-$0.06 $\pm$ 0.21  & $-$0.22 $\pm$ 0.17 & $-$0.11 $\pm$ 0.09 & 0.34
$\pm$ 0.23 & 0.01 $\pm$ 0.11 & $-$0.02 $\pm$ 0.20\\
     &                     &                    &                    &                    &                    & \\

\hline\hline
Star & [Ni/H] & [Rb/H] & [Y/H] & [Ba/H] & $EW$(Li) & A(Li)\\
\hline
51                & $-$0.08 $\pm$ 0.09  &    0.00  &    0.10 $\pm$ 0.29  & 0.46 $\pm$ 0.11 & ---   & ---  \\
52                &    0.04 $\pm$ 0.09  & $-$0.35  &    0.09 $\pm$ 0.51  & 0.82 $\pm$ 0.07 & ---   & ---  \\
\textit{53}$^{*}$ & $-$0.04 $\pm$ 0.10  & $-$0.42  & $-$0.06 $\pm$ 0.17  & 0.00 $\pm$ 0.37 & ---   & ---  \\
\textit{54}$^{*}$ &    0.24 $\pm$ 0.13  &    0.30  &    0.51 $\pm$ 0.08  & 0.74 $\pm$ 0.04 & 244.3 & 1.03 $\pm$ 0.09 \\

\hline
Mean & $-$0.02 $\pm$ 0.09  & $-$0.18 $\pm$ 0.25 & 0.10 $\pm$ 0.20 & 0.72 $\pm$ 0.17 & --- & ---\\
\hline

\end{tabular}

\label{Abund_6664}
\end{center}
\end{table*}

\section{Discussion}

In this work we have observed spectroscopically the largest number of stars in the fields of NGC\,6649 and NGC\,6664 so far. We combined our spectra with archival photometry 
and $Gaia$ DR2 data in order to perform a consistent photometric analysis and properly determine the parameters of each cluster. For the first time we derived stellar atmospheric 
parameters for both blue and evolved members as well as the chemical abundances for the latter (but only for NGC\,6649 and NGC\,6664 stars).

\subsection{Cluster parameters}\label{sec_clust_param}

In this research we have conducted a comprehensive analysis of these three clusters for which we have derived the properties summarised in Table\,\ref{resumen_cum}.
We find that the distance of the clusters derived from the photometric analysis ($d_{\textrm{ZAMS}}$ in Table~\ref{resumen_cum}) is consistent with the one corresponding to their $Gaia$ DR2 parallaxes ($d_{\textrm{GDR2}}$). 
We conclude that these three clusters are coeval, younger than the Pleaides, with an age $\approx$70\,Ma, which is in good agreement with the earliest
blue members found, with spectral types around B5--B6. The age of Be\,55 (63$\pm$15\,Ma) agree with the result of \citet[][30--100\,Ma]{Molina18} and 
it is compatible within the errors with \citet[][50$\pm$10\,Ma]{Be55}.
In addition, our value is also consistent with that of the Cepheid obtained from a different approach \citep[][63$^{+12}_{-11}$\,Ma]{Lohr18}.
According to the isochrones, at the age of these clusters, the mass of the (super)giants is supposed to be $\approx$6\,M$_{\sun}$, with an uncertainty of around 
0.3\,M$_{\sun}$. This value evaluates the difference of mass found when propagating the uncertainty associated with the choice of the isochrone.

\begin{table*}[ht]    
  \caption{Summary of the parameters derived in this work for the clusters under study.\label{resumen_cum}}
\begin{center}
\begin{tabular}{lccccccccc}
\hline\hline
\multirow{2}{*}{Cluster} & $E(B-V)$  &  $d_{\textrm{ZAMS}}$ &  $d_{\textrm{GDR2}}$  & $\tau$ &  RV            & [Fe/H] & $r_{\textrm{c}}$ & $r_{\textrm{t}}$  & $M$          \\
                         & (mag)     &  (kpc)               &   (kpc)               &  (Ma)  &  (km\,s$^{-1}$)& (dex) & ($\arcmin$)      & ($\arcmin$)       & (M$_{\sun}$) \\
\hline
NGC\,6649 & 1.39$\pm$0.06 & 1.70$\pm$0.12 & 2.0$\pm$0.4 & 63$\pm$15 &  $-$8.66$\pm$1.03 & $+$0.02$\pm$0.07 & 1.67$\pm$0.03 & 14.7$\pm$1.4 & $\approx$2\,600 \\
NGC\,6664 & 0.77$\pm$0.05 & 1.78$\pm$0.12 & 2.0$\pm$0.2 & 79$\pm$18 & $+$19.34$\pm$0.21 & $-$0.04$\pm$0.10 & 5.08$\pm$0.41 & 23.0$\pm$5.0 & $\approx$2\,900 \\
Be\,55    & 1.81$\pm$0.15 & 3.24$\pm$0.22 & 3.0$\pm$0.8 & 63$\pm$15 & $-$31.68$\pm$7.37 & $+$0.07$\pm$0.12 & 1.28$\pm$0.06 &  5.4$\pm$0.4 & $\approx$1\,300 \\
     
\hline 
\end{tabular}
\end{center}

\end{table*}

As previously commented (Sect.\,\ref{sec_vrad}) our RVs are consistent with those of \citet{Me08}. The average RVs calculated for NGC\,6664 and Be\,55 are fully compatible with those expected, for their distances along the line-of-sight direction, according to the Galactic rotation curve \citep{Reid14}.
However, this is not true for NGC\,6649, which despite being very close to NGC\,6664 in the sky (separated by only 2.3$^{\circ}$) and at similar distances ($\approx$\,2\,kpc), shows a RV quite different
($-8.7$ and $+19.3$\,km\,s$^{-1}$, respectively). Therefore, NGC\,6649 seems to have a peculiar RV, which is unexpected since it is young enough to be a good tracer of its bith place.
Recently, \citet{Soubiran18} based on $Gaia$ DR2  RVs have recalculated the RVs for hundreds of open clusters, among which are our targets.
For NGC\,6649 they provide a RV=$-$8.87$\pm$0.92\,km\,s$^{-1}$ by using four stars, $-$4.44 (NGC\,6664, only one star) and $-$43.58$\pm$9.24 (Be\,55, four).
These values are merely indicative, since the same stars have not always been considered members by the different authors, becoming their comparison less reliable. For NGC\,6649 the three values \citep[also including the 
result of][]{Me08} show an excellent agreement. In the case of NGC\,6664, \citet{Soubiran18}
estimated a value smaller than ours, which approximates it to that of NGC\,6649 (using only one star). In addition, for Be\,55 their value is compatible, within the errors,
with ours, because the dispersion is very large, although our value (derived from a larger sample) is smaller.
Although the potential of $Gaia$ is huge in this field, at present, the current DR2 still does not provide enough values to properly sample open clusters.
In the next years, upcoming DRs will shed light on this topic, also allowing a better understanding on the Galactic dynamics.

From a simple visual inspection of the sky region around the nominal centre of each cluster we can already have a first impression of their relative appearance.
In an increasing order of size we find Be\,55 (with a diameter around 2--3$\arcmin$), NGC\,6649 ($\approx$5$\arcmin$) and NGC\,6664 (a more diffuse cluster that 
seems to extend up to 7--10$\arcmin$). This trend is shared by the different studies found in the literature. Beyond this fact, results from different papers are not 
directly comparable since the variables computed to characterise the cluster size are somewhat different in each case. In Table~\ref{tab_rad} our results are compared 
to those obtained in recent large surveys. \citet{kha13}, from PPMXL and 2MASS data, obtained angular and physical values for both the core ($r_{\textrm{0}}$ and
$r_{\textrm{c}}$) and limiting radii ($r_{\textrm{2}}$ and $r_{\textrm{t}}$). Since the distances for each cluster (ours and theirs) are different, it is more adequate to 
compare angular values. Results are similar for NGC\,6649 and Be\,55 but not in the case of NGC\,6664, for which our values are significantly larger (since it does not 
stand out from the field). \citet[][Sam17 in Table~\ref{tab_rad}]{sampedro17}, based on astrometry from the UCAC4 catalogue, yielded numbers for the radius very similar 
to those of \citet[][CG18]{cantat18}. This latter, by analysing $Gaia$ DR2 data, provided the radius ($r_{\textrm{50}}$) inside which half of the members identified are contained. The
values of these authors, as expected, are comprised between the core and tidal radii calculated by us. Finally, \citet{Maciejewski07} investigated Be\,55 finding a very 
compact cluster ($r_{\textrm{c}}$=0.7$\pm$0.1\,$\arcmin$) with a limiting radius of 6$\arcmin$.

\begin{table}[ht]
  \caption{Comparison of the cluster sizes (arcmin) derived in this work and in the literature.\label{tab_rad}}
\begin{center}
\begin{tabular}{lcccccc}
\hline\hline
\multirow{2}{*}{Cluster} & \multicolumn{2}{c}{Kha13} & Sam17 & CG18 & \multicolumn{2}{c}{This work}\\
    &  $r_{\textrm{0}}$ & $r_{\textrm{2}}$  &  $r$  & $r_{\textrm{50}}$  & $r_{\textrm{c}}$ & $r_{\textrm{t}}$  \\
\hline
NGC\,6649 &  2.1  & 10.8  &    4.0  &  3.4  &  1.7 & 14.7  \\  
NGC\,6664 &  0.6  &  8.4  &    6.5  &  6.1  &  5.1 & 23.0  \\
Be\,55    &  0.9  &  6.0  &    2.2  &  1.9  &  1.3 &  5.4  \\
     
\hline 
\end{tabular}
\end{center}

\end{table}

Regarding the mass of these clusters, in spite of our rough estimates, we find that NGC\,6649 and NGC\,6664 are moderately massive clusters with
initial masses around 2\,500--3\,000\,M$_{\sun}$ hosting 3--4 (super)giants stars, in good
agreement with that expected from simulations \citep[see discussion in][]{be51}.
However, among the three clusters investigated in this paper, Be\,55 hosts the largest number of evolved stars, despite being the smallest and least massive cluster. 
This evidence could be due to the fact that the lower brightness of the stars in this cluster (roughly 2.5 magnitudes fainter than those in  NGC\,6649
NGC\,6664) has made it difficult to select the members, leaving out a (large?) number of potential candidates.
Our value for its present mass, $\approx$900\,M$_{\sun}$, is slightly larger than that obtained by \citet[][795\,M$_{\sun}$]{Maciejewski07}.
In this age range open clusters containing a similar number of (super)giants are rare (see Table~\,\ref{RG_OC}) as well as significantly more massive. Representative 
examples of this kind of clusters are NGC\,6067 \citep{6067}, Be\,51 \citep{be51} or NGC\,2345 \citep{2345}.

\subsection{Stellar parameters and abundances}

\begin{figure}[ht] 
  \centering         
  \includegraphics[width=\columnwidth]{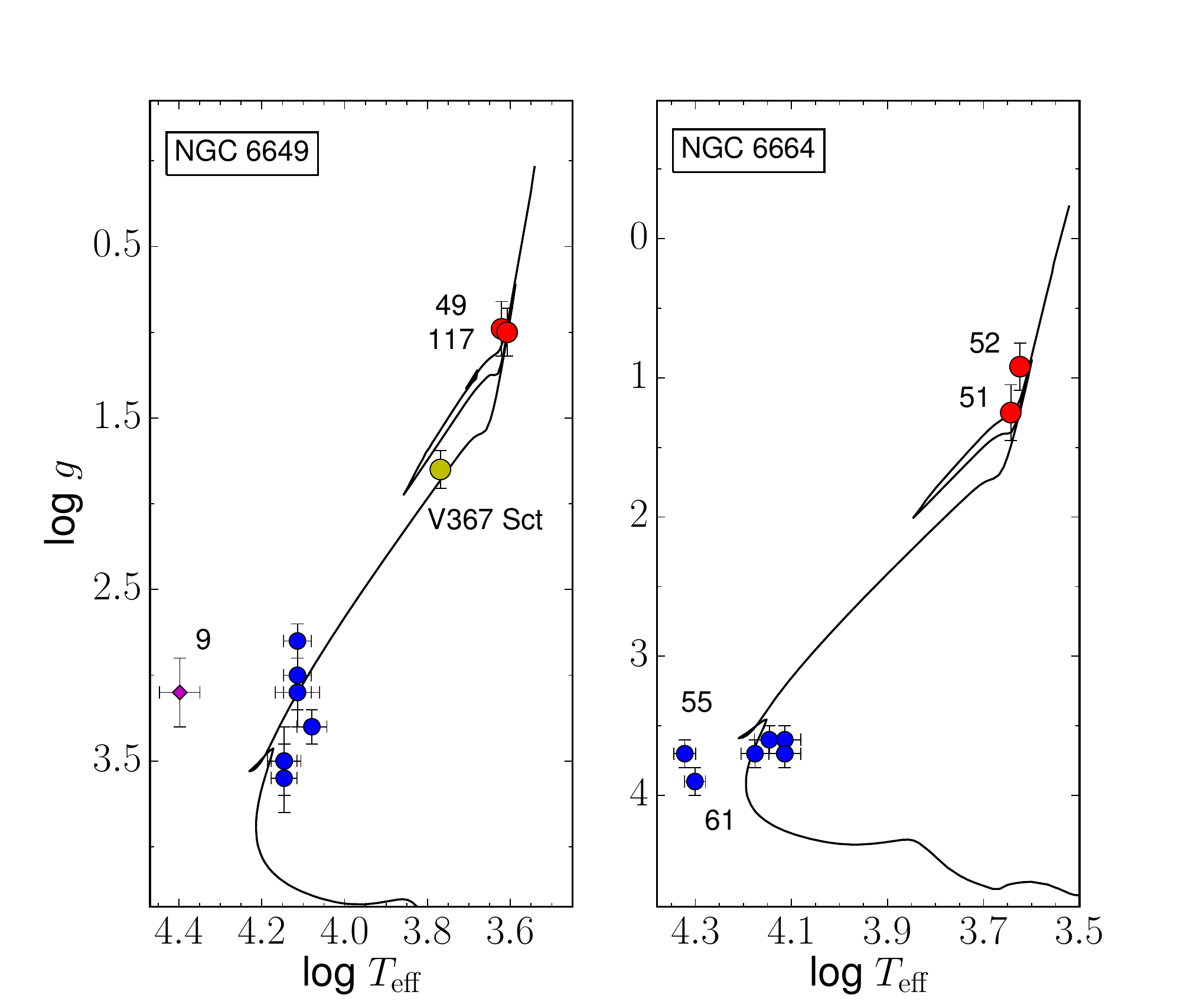}   
  \caption{Kiel diagram for likely members in NGC\,6649 and NGC\,6664. Colours and symbols are the same as those in Fig.~\ref{isoc_6649}.} 
  \label{pHR}
\end{figure}

The reliability of our methodology is corroborated when plotting the Kiel diagram, i.e. log\,$g$/$T_{\textrm{eff}}$ (Fig.~\ref{pHR}). On this diagram, which is 
independent of distance, we can appreciate how well the data derived photometrically such as the isochrone fit the 
spectroscopic results, calculated with two different procedures, for blue and cool stars, respectively. The fitting is better for the latter group
since, unlike the former one, it was observed at high resolution. In any case, with the only exception of the BSSs, as expected, members lie on the isochrones.

\begin{figure}  
  \centering         

  \includegraphics[width=\columnwidth]{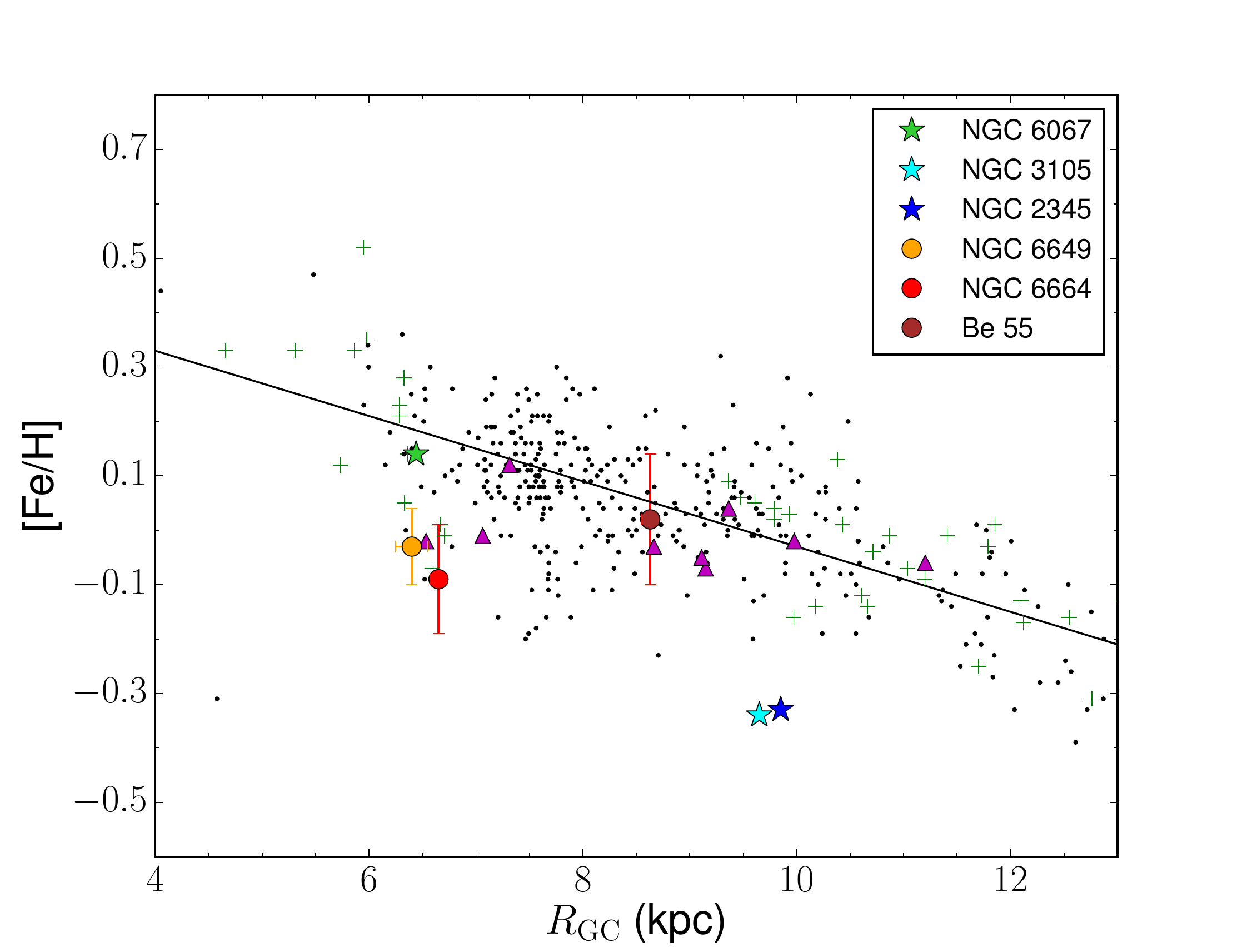}  
  \caption{Iron abundance gradient in the Milky Way found by \citet{Ge13,Ge14}. The black line is the Galactic gradient, green crosses are Cepheids studied in those papers,
  whereas black dots show data for other Cepheids from literature used by these authors. Magenta triangles represent young open clusters in the sample compiled by \citet{net16}. The 
  orange circle is NGC\,6649 and the red one is NGC\,6664. Finally, other clusters analysed by our group with the same technique are marked with star symbols.
  All the values shown in this plot are rescaled to \citet{Ge14}, i.e. R$_{\sun}$=7.95 kpc and A(Fe)=7.50.} 
  \label{gradient_OC} 
\end{figure}

We obtained a solar composition for NGC\,6649 ([Fe/H]$=+0.02\pm0.07$) and NGC\,6664 ([Fe/H]$=-0.04\pm0.10$). As mentioned above, these clusters had not been 
observed spectroscopically at high resolution to date, reason why there are no previous studies with which to compare our results. Instead, we resorted to the 
Galactic gradient and trends. Regarding the former, we took as a reference the work by \citet{Ge13,Ge14}. They estimated the radial distribution of metallicity in the Milky Way ($-$0.06~dex/kpc) by
employing Cepheids. They are very appropriate to contrast our results since Cepheids are young enough ($\tau\approx20$\,--\,400~Ma) to trace present-day abundances.
In Fig.~\ref{gradient_OC}, we display the positions of our clusters on this gradient, also including Be\,55 (represented by its Cepheid, whose metallicity is [Fe/H]$=+0.07\pm0.12$). 
For comparison, we also overplotted some young clusters (i.e. age below 500 Ma) from the sample studied by \citet{net16}
as well as some other young clusters investigated by our group.
Our clusters lie below this gradient, indicating that their metallicities are lower than those expected according to their Galactocentric positions.

\begin{figure*}  
  \centering         
  \includegraphics[width=19cm]{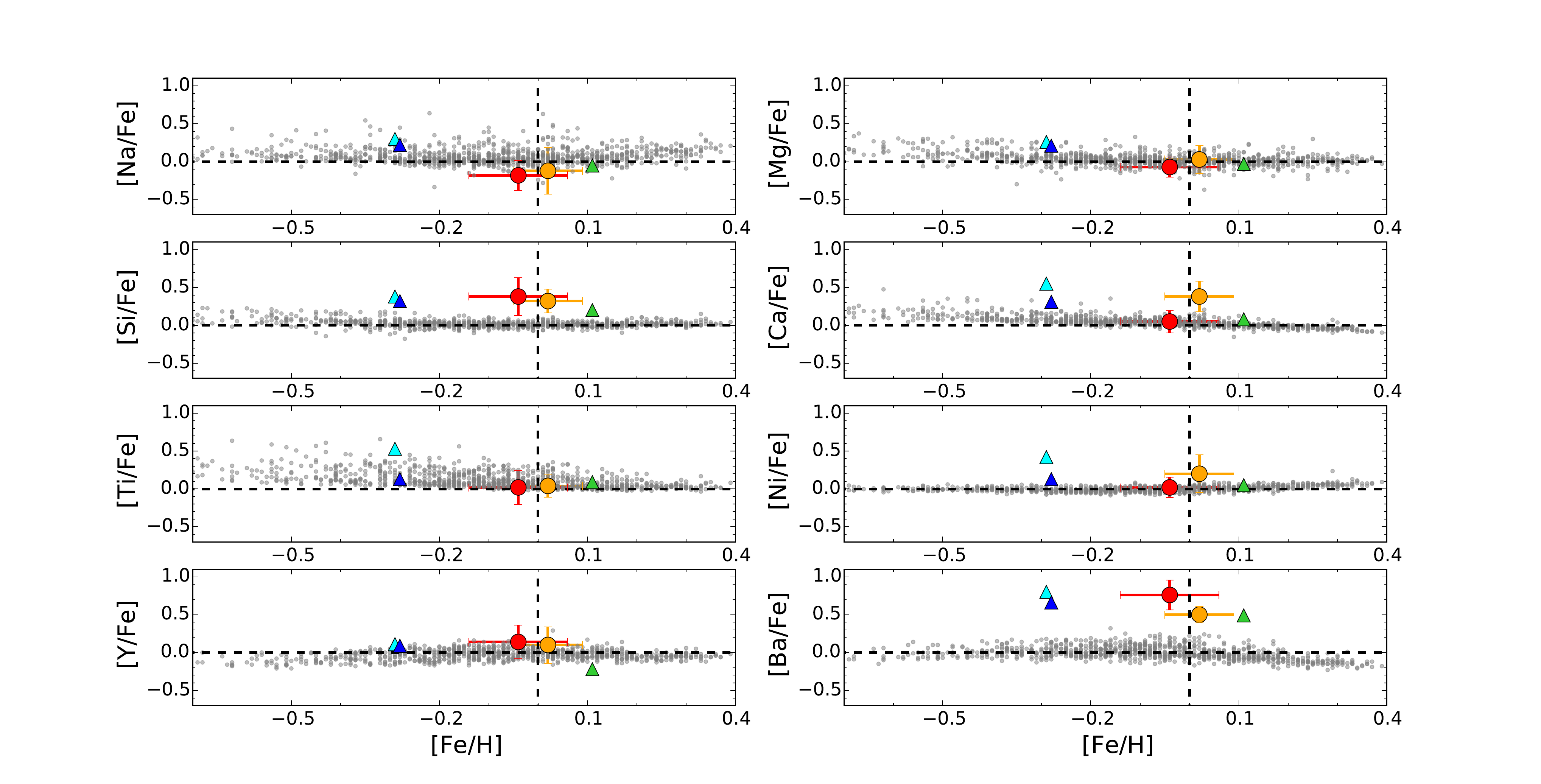}   
  \caption{Abundance ratios [X/Fe] versus [Fe/H]. The grey dots represent the Galactic trends for the thin disc \citep{vardan,elisa17}. NGC\,6067, NGC\,3105 and NGC\,2345 are drawn with triangles (green, cyan and blue, respectively)  
  whereas NGC\,6649 and NGC\,6664 are the orange and red circles, respectively. Clusters are represented by their mean values. The dashed lines indicate the solar value.} 
  \label{trends_OC} 
\end{figure*}

Finally, we also compare our abundances with the Galactic trends for the thin disc (Fig.~\ref{trends_OC}). We plot abundance ratios [X/Fe] vs [Fe/H] obtained by \citet{vardan} for Na, Mg, Si, Ca, Ti and Ni and by \citet{elisa17} for Y and Ba in the workframe 
of the HARPS GTO planet search program. The chemical composition of both clusters is compatible, within the errors, with the Galactic trends observed in the thin disc of the Galaxy.
The [Si/Fe] ratio for both clusters and the [Ca/Fe] for NGC\,6649 lie slightly above the trend, but still consistent with it. The [$\alpha$/Fe] ratios are slightly enhanced, 
presenting values of +0.19 and +0.10 for NGC\,6649 and NGC\,6664, respectively.
We derive a roughly solar [Y/Fe] against a supersolar [Ba/Fe], which is in good agreement with the dependence on age and Galactic location found by \citet{Mi13} by comparing the
abundances of Y and Ba in different open clusters. 
Remarkably, we find a strong over-abundance of barium ([Ba/Fe]$\approx$+0.8), which is in line with that described by \citet{dor09}. 
They studied the evolution of barium with time for dwarf stars in open clusters. They found (see their figure~2) the highest abundances, similar to ours, 
in the youngest clusters of their sample (specially for those clusters with ages $\tau\leq$100\,Ma). These abundances are higher than those predicted by standard
theoretical models. To explain this enrichment of Ba, \citet{dor09} suggested the so-called ``enhanced $s$-process'': 
the use in chemical evolution models of a higher yield of Ba with respect the $s$-process (the contribution of the $r$-process is not so important and its yields are only affected at low metallicity).
Finally, none of these stars have high abundances of Li and Rb as we could expect for a super-AGB, confirming their evolutionary status, massive red giants, inferred from 
their positions on the CMDs. 

\subsection{Blue straggler stars}\label{sec_BSS}

The existence of BSSs in open clusters, and their relationship with the Be phenomenon, has been known for a long time \citep[see][and the references therein]{Me82}.
These stars, because of their anomalous positions on the CMDs have an appearance younger than the rest of the cluster members. They seem to be brighter and bluer 
than the MS stars at the cluster turn-off point. Although several mechanisms have been suggested to explain the origin of the BSSs, in young open clusters the leading
scenario is the mass transfer between companions in a multiple system \citep{MCCrea64,Perets}.

The presence of this kind of objects is very common at this age in Galactic open clusters as we have previously studied \citep[see for example the discussions in][]{amparo_6649,6067}.
In the field of the clusters investigated in this paper, five objects have been proposed as BSS candidates: S9 \citep{amparo_6649} and S35 \citep{Ah95} in NGC\,6649 and S55, S59 and S61 in NGC\,6664 \citep{scm6664,Ah95}.
Based on astrometry (see Table~\ref{tab_mp}), the cluster membership for S59 is discarded. Among the remaining four, when examining our data, only S35 seems to be a normal B-type MS star. On 
the contrary, S9, S55 and S61 fulfill the observational conditions expected for BSS candidates. They have the earliest spectral types in each cluster and occupy peculiar positions on both CMD and 
Kiel diagrams. In addition, here we presented two new mild candidates: S28 (NGC\,6649) and 
S7 (Be\,55). However, regarding the Kiel diagram, S28 occupies a normal position and for S7 atmospheric parameters have not been calculated.
Remarkably, as in the case for example of stars 267 and HD\,145304 in NGC\,6067 \citep{6067}, the spectrum of S55 (BN2\,IV) shows the enhancement of some nitrogen features, which is in good agreement with anomalies 
in the CNO abundances as predicted by the mass-transfer mechanism \citep{Sa96}. Morever, three of the five candidates, S9 (which is even an X-ray source), S61 and S28, are known Be stars which also supports this scenario.

\subsection{Cepheids}

Cepheids are surely the best known variable stars for being primary standard candles. Since the beginning of the 20th century, is well known that their 
brightness is directly proportional to their period \citep{lea12}. However, Cepheids, beyond cosmological or extragalactic implications, are astropysical laboratories 
of great importance in the study of stellar evolution. Cepheids are YSGs (i.e. typically from mid-F to early-G spectral types) with masses in the range 3--10\,M$_{\sun}$, whose progenitors are B-type stars of the MS. Yellow supergiants become
unstable and begin to pulsate, turning into Cepheids, when crossing the instability strip (IS). Cepheids allow us to constrain evolutionary models for intermediate-mass stars, especially when they are hosted in 
open clusters, such as $\delta$\,Cep, the prototype of this type of objects, in the young stellar association of Cep\,OB6 \citep{majaess12}. However, this is an unsual occurrence: despite the fact that more than 3\,000 open 
clusters are known in the Galaxy \citep{khar16}, only 31 contain Cepheids \citep{Chen17}. 
Among them, in this paper we have focused on three targets, whose main properties are shown in Table\,\ref{resumen_cep}. 
In order to minimise the possible effect that the parallax zero-point offset of the current $Gaia$ DR2 \citep{Groenewegen18} may have on our results, the distances listed below 
are calculated as the weighted average of the values derived from the individual parallax for each Cepheid (see Table\,\ref{tab_mp}) and the cluster mean (via ZAMS fitting, Table\,\ref{resumen_cum}).
Tabulated reddenings, $E(B-V)$, are also the average of the cluster mean and the individual values (i. e. 1.33 and 0.62 for V367\,Sct and EV\,Sct, respectively) derived from their average $BV$ magnitudes and the 
calibration of \citet{Fi70}. Finally, for the little-studied Cepheid S5 we have estimated the $JHK_{\textrm{s}}$ average values by employing the 
PLR obtained by \citet{Chen17}.

\begin{table*}[ht]    
  \caption{Summary of the parameters for the Cepheids studied in this work.\label{resumen_cep}}
\begin{center}
\begin{tabular}{lcccccccccc}
\hline\hline
Cepheid   & $P$ (d)   &  $d$ (kpc)  &  $E(B-V)$   &  [Fe/H]       &  $<B>$ & $<V>$   & $<I_{\textrm{c}}>$  &  $<J>$ &  $<H>$  & $<K_{\textrm{s}}>$ \\
\hline
V367\,Sct & 6.29$^a$  &  1.79$\pm$0.29  & 1.36$\pm$0.06  &    0.00$\pm$0.04     & 13.396$^a$ & 11.610$^a$ &  ---       & 7.657$^b$ & 6.983$^b$  & 6.671$^b$ \\ 
EV\,Sct   & 3.09$^a$  &  1.81$\pm$0.23  & 0.70$\pm$0.05  & $-$0.04$\pm$0.10$^c$ & 11.292$^d$ & 10.135$^d$ & 8.668$^d$  & 7.612$^d$ & 7.192$^d$  & 7.015$^d$ \\ 
S5        & 5.85$^e$  &  3.03$\pm$0.37  & 1.81$\pm$0.15  &    0.07$\pm$0.12$^e$ & ---       &  13.834$^e$ &  ---       & 7.987     & 7.686      & 7.608     \\ 
     
\hline 
\end{tabular}
\end{center}
\begin{list}{}{}
\item[]References: $^{a}$ \citet{Fernie95}; $^{b}$ \citet{Chen17}; $^{c}$ Average value derived in this work for its parent cluster; $^{d}$ \citet{Ngeow12}; $^{e}$ \citet{Lohr18}. 
  \end{list}
\end{table*}

\subsubsection{V367\,Sct}

We have only calculated parameters and abundances for one of the three Cepheids aforementioned (V367\,Sct). 
Unfortunately, we have been unable to disentangle the properties of EV\,Sct with our current methodology due to the ambiguous profile 
of the lines displayed in its spectrum.
In the case of the third Cepheid, S5, it could not be analysed because it was observed at lower resolution.

\citet{Ge14} derived the atmospheric parameters for a large number of Cepheids among which V367\,Sct was included. Their results, compared to ours, are 
displayed in Table~\ref{cef_param}. Their errors represent the standard deviation of the different values derived from individual spectra. 
Since Cepheids are pulsating variables their parameters depend on the moment of the observation (especially the effective temperature), reason for which 
the aim of this comparison is only qualitative. Even so, effective temperature and metallicity are compatible within the errors. However, microturbulent 
velocity and surface gravity are rather different. As both works used the same linelist and MARCS models, this difference might reside in the methodology employed. 
They computed the stellar parameters as decoupled quantities whereas we derived them as coupled variables at the same time. According to \citet{torres12} 
this different approach to atmospheric parameters could biase the results.

The chemical composition of the Cepheid is mostly compatible within the errors with those of the parent cluster. The only exceptions are the abundances of
Na and Ca. \citet{Ge15} computed abundances for $\alpha$-elements based on stellar parameters previously obtained in \citet{Ge14}. In Table~\ref{cef_param} 
we also compare their chemical abundances with ours. Both sets of abundances, including Na and Ca, are fully compatible. 
We note that for the Cepheids in common with \citet[][both here and in previous works]{Ge14} our results are compatible within the 
errors with theirs. In our case, the cluster abundances are derived from red (super)giants whereas they used Cepheids for the
Galactic gradient. This is a possible explanation for the offset observed in Fig.\ref{gradient_OC}.

\begin{table}
  \caption{Comparison of the results obtained for the Cepheid V367\,Sct from this work (TW) and \citet{Ge14}, for the stellar atmospheric parameters, and 
  \citet{Ge15},for chemical abundances, relative to solar values by \citet{Gre07}.\label{cef_param}}
\begin{center}
\begin{tabular}{lcc}
\hline\hline
                        &   TW           &  Ge14          \\
\hline
$T_{\textrm{eff}}$\,(K) &  5\,875$\pm$57 & 5\,585$\pm$237 \\
log\,$g$                &  1.80$\pm$0.11 & 1.13$\pm$0.15  \\
$\xi$\,(km\,s$^{-1}$)   &  2.71          & 3.78$\pm$0.33  \\
$[$Fe/H$]$              &  0.00$\pm$0.04 & 0.10$\pm$0.08  \\
                        &                &                \\ 

\hline\hline                        
                        &   TW           &  Ge15          \\
\hline
$[$Na/H$]$              & 0.51$\pm$0.07  & 0.46$\pm$0.04  \\ 
$[$Mg/H$]$              & 0.17$\pm$0.46  & 0.04$\pm$0.14  \\ 
$[$Si/H$]$              & 0.18$\pm$0.12  & 0.20$\pm$0.18  \\
$[$Ca/H$]$              & 0.06$\pm$0.06  & 0.06$\pm$0.12  \\

\hline 
\end{tabular}
\end{center}
\end{table}

\subsubsection{Ages and Masses: pulsating versus evolutionary values}

Cluster Cepheids help us determine the cluster age when fitting isochrones, since they provide an extra anchor point between the brightest blue members and the RSGs. 
The age thus obtained for the cluster, and therefore also for the Cepheid itself, can be checked with that inferred by using a period-age relation (PAR). 
On the one hand, a family of PARs based on theoretical models of stellar evolution yields the pulsating age of the Cepheid from its observed period. Results depend strongly on the input physics. In order to illustrate 
it, we employed a canonical PAR from \citet{bon05}, which does not account for rotation, and a non-canonical one from \citet{An16}, which assumes an average initial rotation ($\omega$=0.5).
On the other hand, there also exists a branch of empirical PARs calibrated with Cepheids hosted in open clusters, such as those of \citet{efremov} and \citet{Turner12}. The former employed cluster Cepheids in the LMC while
the latter focused on Galactic ones. Similarly, we used a mass-luminosity relation \citep[MLR, derived by][]{and14} to estimate the pulsating masses of the Cepheids from their average $V$ magnitudes ($<V>$).
Luminosities were calculated assuming individual distances (Table~\ref{resumen_cep}), bolometric corrections for supergiants given in \citet{Hump84} and an absolute bolometric
magnitude for the Sun, $M_{\textrm{bol}}^{\sun}$=4.74. In Table~\ref{cef_age} we compare the pulsating ages and masses calculated in this way with those values 
inferred from the isochrones.

\begin{table*}[ht]
  \caption{Age\,(Ma) and mass (M$_{\sun}$) of cluster Cepheids derived from their parent cluster (inferred from isochrones) and by using different period-age (PAR) and mass-luminosity (MLR)
  relations.\label{cef_age}}
\begin{center}
\begin{tabular}{lccccccccc}
\hline\hline
\multirow{2}{*}{Cepheid} & \multirow{2}{*}{Cluster} & \multirow{2}{*}{$P$\,(d)} &  \multicolumn{2}{c}{Isochrones} & \multicolumn{4}{c}{PAR$^{*}$ (Age)} & MLR$^{**}$ (Mass) \\
    &      &      &   Age & Mass  & Ef03 & Bo05 & Tu12 & An16 & An14  \\
\hline
V367\,Sct & NGC\,6649 & 6.29  &  63$\pm$15 & 6.2$\pm$0.3 &  96  &  60  &  80  &  101  & 5.9  \\
EV\,Sct   & NGC\,6664 & 3.09  &  79$\pm$18 & 5.6$\pm$0.3 & 152  &  77  & 133  &  153  & 5.1  \\             
S5        & Be\,55    & 5.85  &  63$\pm$15 & 6.2$\pm$0.3 & 100  &  63  &  84  &  105  & 6.4  \\            
     
\hline 
\end{tabular}
\end{center}
\begin{list}{}{}
\item[]$^{*}$ Results derived from period-age relations of \citet[][Ef03]{efremov}, \citet[][Bo05]{bon05}, \citet[][Tu12]{Turner12} and \citet[][An16]{An16}.
\item[]$^{**}$ Mass-luminosity relation of \citet[][An14]{and14}.
  \end{list}
\end{table*}

The differences in the results from both model-based PARs cannot be attributed exclusively to the effects of rotation. Although both PARs are computed for fundamental and first overtone modes Cepheids at the Galactic 
metallicity ($Z_{\textrm{MW}}$), they use different values for it: $Z_{\textrm{MW}}$=0.02 (canonical) and $Z_{\textrm{MW}}$=0.014 (non canonical model). In addition, the first one takes into 
account all the evolutionary phases inside the IS whereas the non-canonical PAR is only valid for second and third IS crossings. Evolutionary ages derived from isochrones are in excellent agreement with
the canonical PAR of \citet{bon05}, since the rotating PAR gives older (unrealistic) ages, not compatible with the stellar B-type population observed in these clusters.
Concerning the empirical PARs, which are not especific for the pulsation mode or the IS crossing number, they yield older ages than that of \citet{bon05}.
The PAR of \citet{Turner12} gives realistic numbers (with the exception of EV\,Sct), whereas the estimates obtained from the relation of \citet{efremov} are very similar to those 
derived from the rotating PAR.
Regarding the mass, although a mass discrepancy would be expected \citep{and14}, both approaches provide a similar value.

\subsubsection{YSG/RSG ratio}

Throughout the life of Cepheids, IS crossing occurs up to three occasions\footnote{According to some evolutionary models \citep[see e.g.][for references]{Turner06} in some cases a fourth and fifth
crossings could occur during the He-shell burning phase, but the reality of these extra crossings has not been confirmed \citep{An16}.}: the first one before the first dredge-up (in the H-shell burning phase) and the other two, during the blue loop 
(He-core burning phase), where they spend most of their lifetimes as Cepheids \citep{and14}. 
The Cepheids/RGs ratio observed is very sensitive to the extension of the blue loop, which strongly depends on metallicity and the input physics assumed
by the theoretical models \citep{mat82,eks12,and14,wal15}.
According to this, in order to check theoretical predictions with observations, in a previous paper we looked in the literature for young open clusters that host Cepheids. 
As we only found a dozen, we decided to enlarge our sample, to make it more representative, including all the clusters with ages between 50 and 100\,Ma observed 
by \citet{Me08} and counting all the YSGs, not only Cepheids \citep[the reader can find a detailed description in][]{6067}. In this way we calculated a 
yellow-to-red (super)giants ratio, YSG/RSG=0.19. According to \citet{eks12}, at solar metallicity, we estimated that this ratio was expected to be around, YSG/RSG$\approx$0.6 
(considering models with and without rotation), in good agreement with \citet{mat82}, in the case in which they included a moderate overshooting in the models. 
Therefore, the observed value was lower than those predicted by stellar evolutionary models.

As seen before (Sect.~\ref{membership}), in light of $Gaia$ DR2 astrometry we have had to discard the membership of some (super)giants which, to date, were assumed to be 
bona-fide members according to their positions on the respective CMDs and their RVs, compatible with the cluster value \citep{Me08}. To check the impact that these 
false members may have had on our previous study, we have reviewed all the clusters in our sample \citep{6067} to consider the membership of each star according to the probabilities 
estimated by \citet{cantat18}. The new values are displayed in Table~\ref{RG_OC}. In total, we examined 54 clusters (with ages in the 50--100\,Ma range) finding 89 RSGs and 
18 YSGs, which represents a YGS/RSG$\approx$0.2, the same result previously obtained, also compatible with younger clusters, such as NGC\,3105 \citep{3105}. 
Consequently, what we previously inferred still remains valid: theoretical models predict a higher ratio YSG/RGS than that observed in young open clusters,
in the domain of intermediate-mass stars.  
Additionally, the disagreement between theoretical predictions and observations also becomes evident from the unexpected positions of the Cepheids on the CMDs. According to 
\citet{and14} almost all Cepheids ($\approx$99$\%$) should be observed during their passage through the blue loop. Nevertheless, it is noticeable that this is not what
we appreciate in these clusters, in which two of the three Cepheids are crossing the Hertzsprung gap.

\begin{table*}[ht]
\caption{Number of red (super)giant stars (and yellow supergiants, Cepheids or not, in brackets) in open clusters with ages between 50 and 150 Ma \citep[from][]{Me08}. 
The age of every cluster has been taken from the WEBDA database.\label{RG_OC}}
\begin{center}
\begin{tabular}{lccclcc}   
\hline\hline
Cluster & $\log\,\tau$ & $N_{\textrm{RSG}}$ & & Cluster & $\log\,\tau$ & $N_{\textrm{RSG}}$\\
\hline
NGC 0129 & 7.90 & 0 (1) &        &  NGC 6087      & 8.00 & 0 (1) \\
NGC 0225 & 8.11 & 0     &        &  NGC 6124      & 8.15 & 7     \\    
NGC 0436 & 7.93 & 1 (1) &        &  NGC 6192      & 8.13 & 5     \\    
NGC 1647 & 8.16 & 2     &        &  NGC 6405      & 7.97 & 1     \\ 
NGC 1778 & 8.16 & 1     &        &  NGC 6416      & 8.09 & 0     \\
NGC 2168 & 7.98 & 2 (1) &        &  NGC 6520      & 7.72 & 2 (1) \\
NGC 2186 & 7.74 & 1     &        &  NGC 6546      & 7.89 & 1     \\
NGC 2232 & 7.73 & 0     &        &  NGC 6649$^{*}$      & 7.80 & 2 (2) \\
NGC 2323 & 8.10 & 0     &        &  NGC 6664$^{*}$      & 7.85 & 2 (1) \\
NGC 2345 & 7.75$^{*}$ & 5     &        &  NGC 6694      & 7.93 & 2     \\
NGC 2354 & 8.13 & 10    &        &  NGC 6709      & 8.18 & 2     \\
NGC 2422 & 7.86 & 0     &        &  NGC 6755      & 7.72 & 3     \\
NGC 2516 & 8.05 & 3     &        &  NGC 7031      & 8.14 & 1     \\
NGC 2546 & 7.87 & 2     &        &  NGC 7063      & 7.98 & 0     \\
NGC 2669 & 7.93 & 1     &        &  NGC 7654      & 7.76 & 0 (1) \\
NGC 2972 & 7.97 & 2     &        &  NGC 7790      & 7.90 & 0 (3) \\
NGC 3033 & 7.85 & 0     &        &  Berkeley 55$^{*}$   & 7.85 & 5 (1) \\
NGC 3114 & 8.09 & 6 (1) &        &  Collinder 258 & 8.03 & 1     \\
NGC 3228 & 7.93 & 0     &        &  IC 2488       & 8.11 & 3     \\
NGC 3247 & 8.08 & 1     &        &  IC 4725       & 7.97 & 1 (1) \\
NGC 4609 & 7.89 & 0     &        &  Melotte 20    & 7.85 & 0     \\
NGC 5138 & 7.99 & 0     &        &  Melotte 101   & 7.89 & 1     \\
NGC 5168 & 8.00 & 0     &        &  Trumpler 2    & 8.17 & 1     \\
NGC 5617 & 7.92 & 3     &        &  Trumpler 3    & 7.83 & 1     \\
NGC 5662 & 7.95 & 2(1)  &        &  Trumpler 9    & 8.00 & 0 (1) \\ 
NGC 5749 & 7.73 & 0     &        &  Trumpler 35$^{**}$   & 7.86 & 0     \\
NGC 6067 & 7.95$^{*,**}$ & 10 (1)&        &  vdBergh 1     & 8.03 & 1 (1) \\
          
\hline
\end{tabular}
\end{center}
\begin{list}{}{}
\item[]$^{*}$ Ages derived by our group in this paper (for NGC\,6649, NGC\,6664 and Be\,55)
  or in previous works \citep[][for NGC\,6067 and NGC\,2345, respectively]{6067,2345}.
\item[]$^{**}$ According to \citet{cantat18} Cepheids QZ\,Nor (NGC\,6067) and RU\,Sct (Trumpler\,35) are not considered cluster 
  members and, therefore, to maintain homogeneity, they have not been taken into account in this table. However,
  the cluster membership of QZ\,Nor should not be questioned as concluded by \citet{Ma13}, who, unlike \citet{cantat18}, conducted
  a comprehensive investigation exclusively devoted to clarify this issue.
  \end{list}
\end{table*}

\section{Conclusions}

We have observed three young open clusters, namely NGC\,6649, NGC\,6664 and Be\,55, which share an unsual occurrence: each of them hosts a Cepheid.
We have collected the largest spectroscopic sample of these clusters to date. For each of the clusters, we compiled a list of preferred members by selecting 
only the high-probability members according to \citet{cantat18}, whose selection was based on $Gaia$ DR2 astrometry. Then, in conjunction with our observations,
we reanalysed archival photometry focusing on our list. As a result of those analyses,
we placed the clusters at distances compatible with those inferred from their $Gaia$ DR2 average parallaxes, slightly below 2\,kpc for NGC\,6649 
and NGC\,6664 and somewhat further 3\,kpc in the case of Be\,55.
We have also determined accurate ages for these clusters, with similar 
values around 70\,Ma, which are in good agreement with the spectral type found for the earliest blue stars, around B5--B6. At this age, the mass of the evolved 
stars are roughly 6\,M$_{\sun}$, in the massive range of the intermediate-mass stars.
We also have computed the size of these clusters, estimating their initial masses in the range 1\,000--3\,000\,M$_{\sun}$. In addition, we confirmed 
the presence of two BSSs in NGC\,6649 and one in NGC\,6664, claiming for two other mild BSS candidates (in NGC\,6649 and Be\,55, respectively). 

$Gaia$ DR2 astrometry, very accurate and available for many stars, could have a certain impact on the study of open clusters.
Upcoming data releases are expected to fix offsets and systematics that allow us to improve our knowledge in this field.
Changes in the membership status of some 
evolved members, that are not very numerous in this age range, could modify cluster parameters such as the age when fitting isochrones. In NGC\,6664 the membership of 
two of the four known RSG members is discarded. However, we find a likely candidate to be a new YSG member, star BD-08\,4641, whose astrometric properties 
as well as its position on the CMDs are fully compatible with cluster membership. Future spectroscopic observations should be of great help to confirm this assumption.

For the first time, atmospheric stellar parameters are computed for stars in NGC\,6649, NGC\,6664 and Be\,55. For the first two clusters we have analysed both 
hot and cool stars, whereas for the third one, Be\,55, we have only studied blue members. The procedure followed in our investigation has proved to be robust
since parameters obtained for both sets of stars using different codes and techniques agree quite well, as shown in the Kiel diagrams. The positions of the stars 
in these diagrams (derived from the spectroscopic analysis) properly lie on the isochrones (which correspond to the photometric ages). 

Also for the first time, we have studied the chemical composition of NGC\,6649 and NGC\,6664 from their evolved stars. We have determined chemical abundances of 
Li, O, Na, $\alpha$-elements (Mg, Si, Ca, and Ti), the Fe-group (Ni), and $s$-elements (Rb, Y, and Ba). Both clusters show a solar composition ([Fe/H]=+0.02$\pm$0.07
and [Fe/H]=$-$0.04$\pm$0.10, respectively) but they are slightly metal-poor relative to the mean of the Galactic gradient. They exhibit a mild $\alpha$-enhancement 
([$\alpha$/Fe]=+0.19 and [$\alpha$/Fe]=+0.10, respectively) and a chemical composition compatible with the trends observed in the Galactic thin disc.
As expected for RGs which have not yet begun the AGB phase, none of the evolved stars show representative abundances of either Li or Rb. However, we have found a significative overabundance 
of Ba ([Ba/Fe]$\approx$0.8\,dex), which supports the theoretical scenario of the enhanced $s$-process.

Cepheids in open clusters offer us the possibility to better calibrate the PLRs for anchoring the distance ladder and thus contribute to mitigating the 
$H_{\textrm{0}}$ tension. They give us the opportunity to test evolutionary models as well. Observables such as the ratio of yellow to red (super)giants (lower than expected)
and the determination of the Cepheid properties (mass, age) from the study of the parent cluster are of great importance to constrain theoretical models. 
The differences detected between observations and predictions reflect the uncertainty that current evolutionary models of the most massive intermediate-mass stars 
still suffer from.

\section*{Acknowledgements}

We thank the anonymous referee for his/her helpful suggestions which have helped to improve this paper.
This research is based on observations collected with the MPG/ESO 2.2-meter Telescope operated at the La Silla Observatory (Chile) jointly by the Max Planck Institute 
for Astronomy and the European Organization for Astronomical Research in the Southern hemisphere under ESO programme 095.A-9020(A).  
This research is also based on observations made with the Mercator telescope (operated by the Flemish Community) and the INT and WHT telescopes (operated by the
Isaac Newton Group of Telescopes) on the islad of La Palma at the Spanish Observatorio del Roque de los Muchachos of the Instituto de Astrof\'{i}sica de Canarias. 
Some observations were obtained with the HERMES spectrograph, which is supported by the Research Foundation -- Flanders (FWO), Belgium, the Research Council of KU Leuven, Belgium, 
the Fonds National de la Recherche Scientifique (F.R.S.--FNRS), Belgium, the Royal Observatory of Belgium, the Observatoire de Gen\`{e}ve, Switzerland and the Th\"{u}ringer Landessternwarte 
Tautenburg, Germany. 
This research is partially supported by the Spanish Ministerio de Ciencia e Innovaci\'{o}n under grants AYA2015-68012-C2-2-P and
PGC2018-093741-B-C21/C22 (MICI/AEI/FEDER, UE).
The authors acknowledge financial support from the FCT - Funda\c{c}{\~a}o para aCi{\^e}ncia e a Tecnologia through national funds (PTDC/FIS-AST/28953/2017)
and by FEDER - Fundo Europeu de Desenvolvimento Regional through COMPETE2020 -- Programa Operacional Competitividade e Internacionaliza\c{c}{\~a}o
(POCI-01-0145-FEDER-028953).
This research has made use of the Simbad database, operated at CDS, Strasbourg (France). This publication also made use of data products from the Two Micron All Sky Survey, 
which is a joint project of the University of Massachusetts and the Infrared Processing and Analysis Center/California Institute of Technology, funded by the National Aeronautics and Space
Administration and the National Science Foundation. 



\bibliographystyle{aa}
\bibliography{latex} 


\appendix

\section{Additional material}\label{appendix}

\begin{table*}
\caption{$Gaia$ DR2 astrometric data and cluster membership for all stars observed spectroscopically in this work.}
\label{tab_mp}
\begin{center}
\begin{tabular}{lcccc}   
\hline\hline
Star    & $\varpi$ (mas)  &  $\mu_{\alpha*}$ (mas\,a$^{-1}$) & $\mu_{\delta}$ (mas\,a$^{-1}$) & Member \\
\hline
\multicolumn{5}{c}{NGC\,6649}\\    
\hline
9     &   0.5423 $\pm$ 0.0551  &  $-$0.085 $\pm$ 0.091  &    0.145 $\pm$ 0.086   &  y \\
14    &   0.4063 $\pm$ 0.0489  &     0.055 $\pm$ 0.074  &    0.016 $\pm$ 0.065   &  y \\  
19    &   0.4514 $\pm$ 0.0582  &     0.057 $\pm$ 0.082  &    0.084 $\pm$ 0.071   &  y \\
23    &   0.4683 $\pm$ 0.0546  &  $-$0.148 $\pm$ 0.081  & $-$0.193 $\pm$ 0.070   &  y \\  
28    &   0.3988 $\pm$ 0.0628  &     0.104 $\pm$ 0.100  & $-$0.155 $\pm$ 0.083   &  y \\
33    &   0.4111 $\pm$ 0.0618  &     0.114 $\pm$ 0.087  & $-$0.084 $\pm$ 0.073   &  y \\
35    &   0.4233 $\pm$ 0.0585  &  $-$0.003 $\pm$ 0.105  & $-$0.147 $\pm$ 0.083   &  y \\
42\,B &   1.6811 $\pm$ 0.0491  &     1.136 $\pm$ 0.075  & $-$7.698 $\pm$ 0.066   &  n \\
42\,K &   1.7601 $\pm$ 0.0791  &     1.641 $\pm$ 0.118  & $-$7.487 $\pm$ 0.102   &  n \\
48    &   0.5130 $\pm$ 0.0505  &  $-$0.104 $\pm$ 0.091  &    0.072 $\pm$ 0.079   &  y \\
49    &   0.4337 $\pm$ 0.0762  &  $-$0.112 $\pm$ 0.124  & $-$0.127 $\pm$ 0.108   &  y \\
52    &   ---                  &  ---                   & ---                    &  y? \\
58    &   0.3636 $\pm$ 0.0552  &  $-$0.021 $\pm$ 0.081  & $-$0.402 $\pm$ 0.070   &  y \\
64    &   0.4203 $\pm$ 0.0528  &     0.066 $\pm$ 0.075  & $-$0.020 $\pm$ 0.065   &  y \\
111   &   0.7927 $\pm$ 0.1924  &  $-$2.063 $\pm$ 0.270  & $-$9.210 $\pm$ 0.226   &  n \\
117   &   0.5630 $\pm$ 0.0859  &  $-$0.147 $\pm$ 0.131  & $-$0.127 $\pm$ 0.115   &  y \\
\hline
\multicolumn{5}{c}{NGC\,6664}\\    
\hline
50    & 1.5688 $\pm$ 0.0632  &     1.683 $\pm$ 0.097  & $-$8.435 $\pm$ 0.106  &  n \\
51    & 0.4640 $\pm$ 0.0826  &  $-$0.074 $\pm$ 0.115  & $-$2.491 $\pm$ 0.105  &  y \\
52    & 0.4603 $\pm$ 0.0601  &     0.240 $\pm$ 0.109  & $-$2.569 $\pm$ 0.092  &  y? \\
53    & 0.4841 $\pm$ 0.0619  &     1.296 $\pm$ 0.112  & $-$0.504 $\pm$ 0.101  &  n \\
54    & 1.0205 $\pm$ 0.0558  &  $-$0.622 $\pm$ 0.085  & $-$2.532 $\pm$ 0.079  &  n \\
55    & 0.3843 $\pm$ 0.0590  &  $-$0.206 $\pm$ 0.102  & $-$2.877 $\pm$ 0.090  &  y \\
56    & 0.3177 $\pm$ 0.0855  &     0.102 $\pm$ 0.119  & $-$3.212 $\pm$ 0.113  &  n \\
59    & 0.3920 $\pm$ 0.0555  &  $-$0.027 $\pm$ 0.087  & $-$1.074 $\pm$ 0.083  &  n \\
60    & ---                  &  ---                   & ---                   &  y? \\
61    & 0.4403 $\pm$ 0.0540  &  $-$0.117 $\pm$ 0.084  & $-$2.496 $\pm$ 0.077  &  y \\
62    & 0.4563 $\pm$ 0.0464  &     0.079 $\pm$ 0.079  & $-$2.207 $\pm$ 0.078  &  y \\
63    & 0.4829 $\pm$ 0.0518  &     0.013 $\pm$ 0.082  & $-$2.661 $\pm$ 0.077  &  y \\
64    & 0.5117 $\pm$ 0.0527  &  $-$0.217 $\pm$ 0.084  & $-$2.693 $\pm$ 0.076  &  y \\
80    & 0.4969 $\pm$ 0.0544  &  $-$0.240 $\pm$ 0.084  & $-$2.592 $\pm$ 0.079  &  y \\
228   & 0.4911 $\pm$ 0.0615  &  $-$0.059 $\pm$ 0.091  & $-$2.849 $\pm$ 0.082  &  y? \\
A     & 0.4873 $\pm$ 0.0680  &  $-$0.043 $\pm$ 0.118  & $-$2.550 $\pm$ 0.099  &  y \\
\hline
\multicolumn{5}{c}{Berkeley\,55}\\    
\hline
1   & 0.3882 $\pm$ 0.0656  &  $-$4.396 $\pm$ 0.152  & $-$4.849 $\pm$ 0.125  &  y \\ 
2   & 0.3804 $\pm$ 0.0466  &  $-$3.865 $\pm$ 0.106  & $-$4.804 $\pm$ 0.085  &  y \\ 
3   & 0.2598 $\pm$ 0.0579  &  $-$4.080 $\pm$ 0.127  & $-$4.752 $\pm$ 0.102  &  y \\ 
4   & 0.3347 $\pm$ 0.0528  &  $-$4.282 $\pm$ 0.103  & $-$4.656 $\pm$ 0.083  &  y \\ 
5   & 0.3484 $\pm$ 0.0416  &  $-$3.875 $\pm$ 0.081  & $-$4.546 $\pm$ 0.065  &  y \\ 
6   & 0.4298 $\pm$ 0.0613  &  $-$4.263 $\pm$ 0.124  & $-$4.803 $\pm$ 0.101  &  y \\ 
7   & 0.3328 $\pm$ 0.0221  &  $-$4.154 $\pm$ 0.042  & $-$4.597 $\pm$ 0.036  &  y \\ 
10  & 0.3562 $\pm$ 0.0316  &  $-$4.211 $\pm$ 0.064  & $-$4.584 $\pm$ 0.054  &  y \\ 
11  & 0.3689 $\pm$ 0.0309  &  $-$3.976 $\pm$ 0.062  & $-$4.546 $\pm$ 0.050  &  y \\ 
12  & 0.3122 $\pm$ 0.0345  &  $-$4.105 $\pm$ 0.073  & $-$4.598 $\pm$ 0.061  &  y \\ 
17  & 0.3218 $\pm$ 0.0385  &  $-$4.015 $\pm$ 0.075  & $-$4.574 $\pm$ 0.061  &  y \\ 
61  & 0.3231 $\pm$ 0.0805  &  $-$2.746 $\pm$ 0.150  & $-$2.613 $\pm$ 0.114  &  n \\ 
\hline
\end{tabular}
\end{center}
\end{table*}


\begin{figure*}[!]  
  \centering         
  \includegraphics[width=13cm]{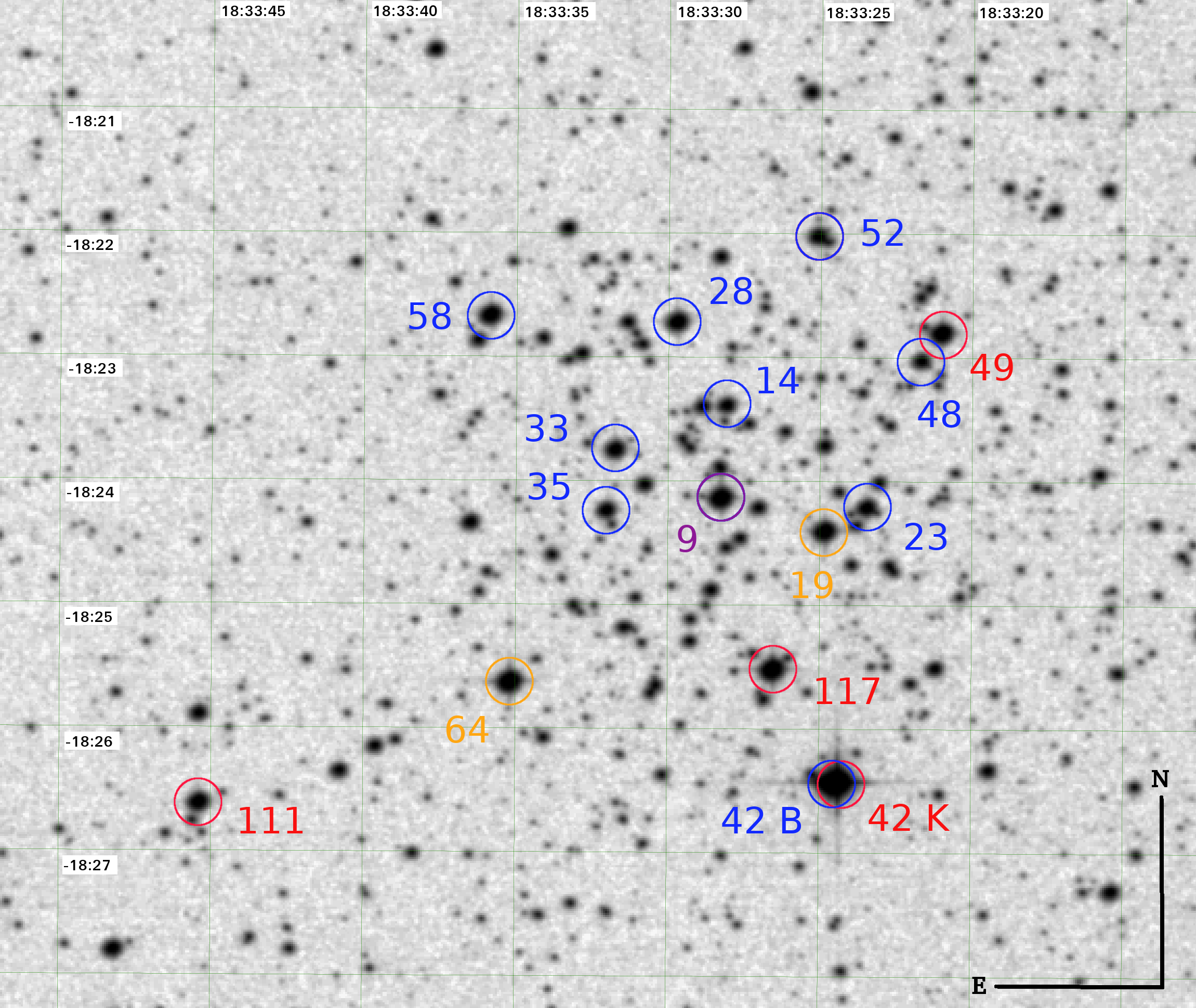}   
  \caption{Finding chart of NGC\,6649 from a 9$'$x9$'$ POSS2 Red image. Stars observed spectroscopically are marked with diferent colours according to their spectral types. 
  The identification of each star corresponds to the WEBDA numbering for this cluster. North is up and East is left.} 
  \label{carta_6649}  
\end{figure*}



\begin{figure*}[!]  
  \centering         
  \includegraphics[width=12cm]{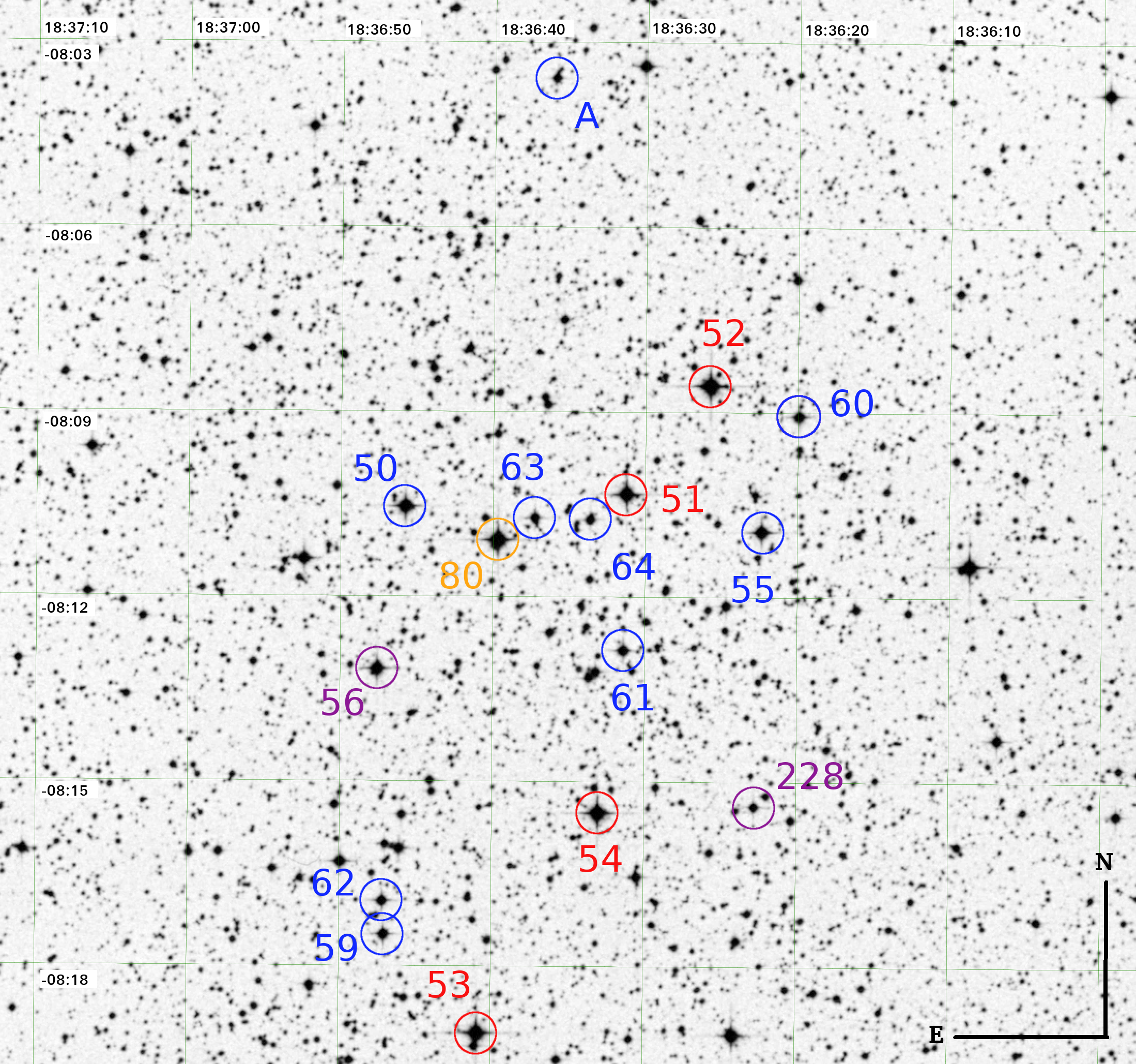}   
  \caption{Finding chart of NGC\,6664 from a 19$'$x17$'$ POSS2 Red image. Stars observed spectroscopically are marked with diferent colours according to their spectral types.   
  The identification of each star corresponds to the WEBDA numbering for this cluster. North is up and East is left.} 
  \label{carta_6664}  
\end{figure*}



\begin{figure*}[!]  
  \centering         
  \includegraphics[width=13cm]{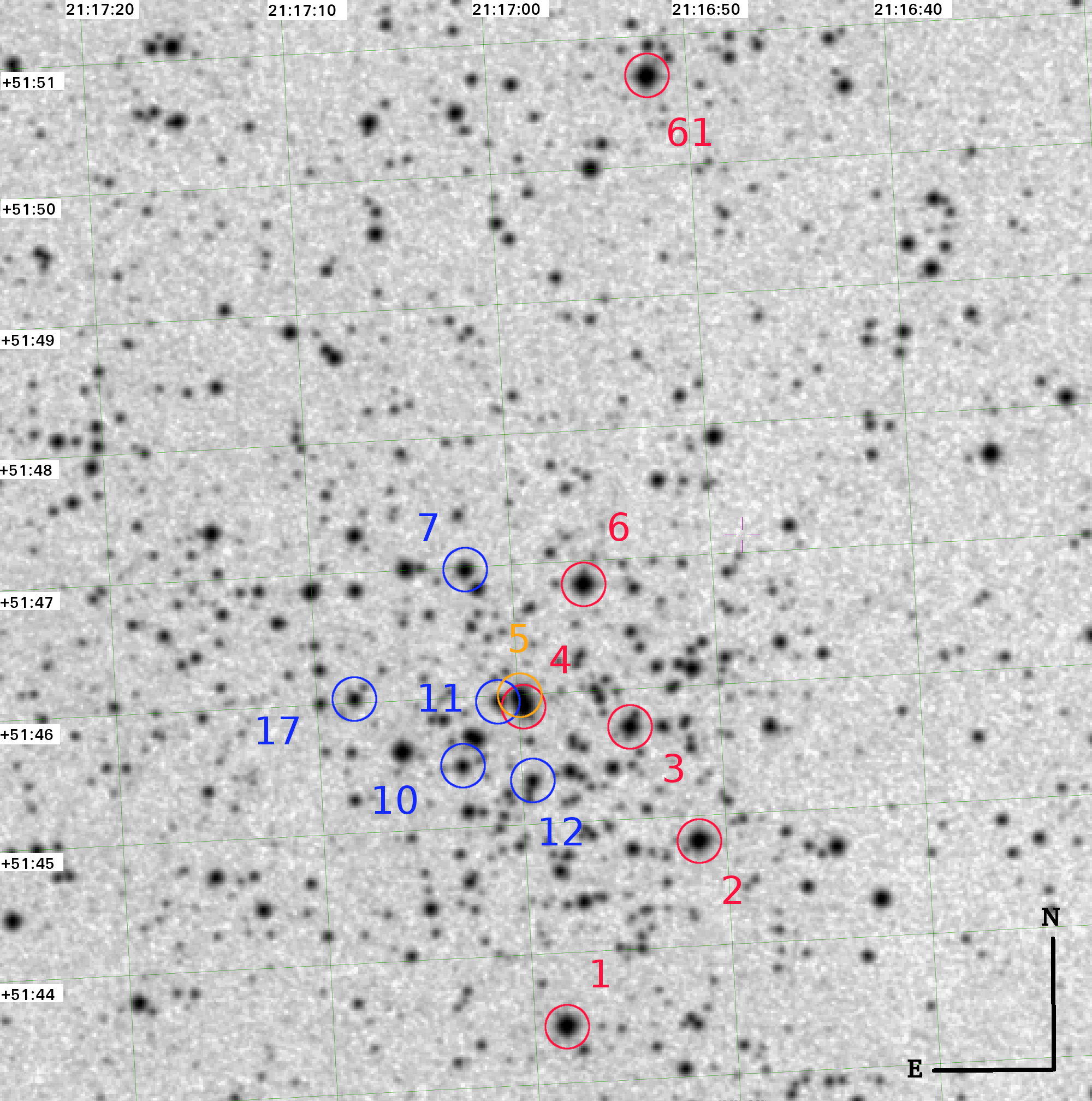}   
  \caption{Finding chart of Be\,55 from a 8$'$x8$'$ POSS2 Red image. Stars observed spectroscopically are marked with diferent colours according to their spectral types.  
  The identification of each star follows the numbering given by \citet{Be55}. North is up and East is left.} 
  \label{carta_be55}  
\end{figure*}



\end{document}